\documentclass[useAMS]{mn2e}
\usepackage{graphicx}

\title[NGC 752, NGC 1817, NGC 2360, NGC 2506]
  {Comprehensive abundance analysis of red giants in the open clusters NGC 752, NGC
1817, NGC 2360 and NGC 2506$^{\star}$ }
\author[A. B. S. Reddy, S. Giridhar and D. L. Lambert ]
  { Arumalla B.S. Reddy$^1$\thanks{E-mail: sudha@iiap.res.in (ABSR);
giridhar@iiap.res.in (SG); dll@astro.as.utexas.edu (DLL)},
    Sunetra Giridhar$^1$ and David L. Lambert$^2$  \\
  $^1$Indian Institute of Astrophysics, Bangalore 560034, India \\
  $^2$W.J. McDonald Observatory, The University of Texas at Austin, Austin, TX 78712
- 0259, USA }
\begin{document}

\pagerange{\pageref{firstpage}--\pageref{lastpage}} \pubyear{2011}

\maketitle

\label{firstpage}

\begin{abstract}
We have analyzed high-dispersion echelle spectra ($R \ga 50,000$) of red giant
members for four open clusters to derive abundances for many elements.
The spread in
temperatures and gravities being
very small among the red giants nearly the same stellar lines were employed
 thereby reducing the random
errors. The errors of average abundance for the cluster were generally in 0.02 to
0.07 dex range. Our present sample covers galactocentric distances  of 8.3 to
10.5 kpc. The
[Fe/H] values are $-$0.02$\pm$0.05 for NGC 752, $-$0.07$\pm$0.06 for NGC
2360, $-$0.11$\pm$0.05 for NGC 1817 and $-$0.19$\pm$0.06 for NGC 2506. 
Abundances relative to Fe for elements from Na to Eu are equal within measurement
uncertainties to published abundances for thin disk giants in the field. This supports the 
view that field stars come from disrupted open clusters.
   
\end{abstract}


\section{Introduction}

Open clusters (OCs) are believed to be 
coeval groups of stars born from the same proto-cluster cloud which may have been
part of a
larger star-forming  region in the Milky Way. Ages of clusters range from very
young where stars are still forming to nearly 10 Gyr (Dias et al. 2002).
Since all stars  in most OCs are at the same distance and have the same chemical
composition, 
basic stellar parameters like age, distance, and metallicity can be determined more
accurately 
than for field stars.  Thus, OCs provide an excellent opportunity to map the structure,
kinematics, and chemistry of the Galactic disk with respect to Galactic coordinates and
time.

The presence of chemical homogeneity among cluster members has been 
shown
by the study of OCs, see, for example, spectroscopic analyses of the Hyades 
(Paulson et al.
2003; De Silva et al. 2006) and Collinder 261 (Carretta et al. 2005; De Silva et al.
2007). 
This observed homogeneity signifies that the proto-cloud is well mixed, 
 and 
hence, the abundance pattern of a cluster bears the signature of chemical evolution
of the natal cloud.
Chemical evolution of the Milky Way has, of course, been well studied using field
stars. A large
fraction of field stars are from disrupted OCs (Lada \& Lada 2003). The youngest OCs
may be intact.
The oldest OCs may be totally disrupted. Thus, the field stars do not fully sample the
age distribution of OCs and, in particular, the youngest stellar generations are
under-represented
by field stars.

In this paper, we report abundance analyses from high-resolution spectra of red
giants 
in four OCs: NGC 752, NGC 1817, NGC 2360, and NGC 2506. These analyses are the
first for these OCs to report elemental abundances for elements from Na to Eu.

The structure of the paper is as follows: 
In Section 2 
we describe the data selection, observations and data reduction and Section 3 is
devoted 
to the abundance analysis. In Section 4 we present our results and compare them 
with the abundances derived from samples of field thin and thick disk
stars (i.e. dwarfs and giants).
 Finally, in Section 5 we give the conclusions.

\begin{table*}
 \centering
  \caption{Target clusters and their properties from the literature.}
\begin{tabular}{lcccccccl}
  \hline
\multicolumn{1}{l}{Cluster}&  
\multicolumn{1}{c}{l}&  
\multicolumn{1}{c}{b}&  
\multicolumn{1}{c}{Age}&  
\multicolumn{1}{c}{[Fe/H]$_{\rm phot.}$}&  
\multicolumn{1}{c}{R$_{gc}$}&
\multicolumn{1}{c}{(m-M)$_{v}$}&
\multicolumn{1}{c}{E(B-V)}&
\multicolumn{1}{l}{[Fe/H]$_{ref}$} \\
\multicolumn{1}{c}{}&
\multicolumn{1}{c}{(deg.)}&
\multicolumn{1}{c}{(deg.)}&
\multicolumn{1}{c}{(Gyr)}&
\multicolumn{1}{c}{(dex.)}&
\multicolumn{1}{c}{(kpc)}&
\multicolumn{1}{c}{(mag.)}&
\multicolumn{1}{c}{(mag.)}&
\multicolumn{1}{l}{ } \\
 \hline

NGC  752 & 137.12 & $-$23.25 & 1.12 & $-$0.12 &  8.3 &  8.18 & 0.03 &
Barta\v{s}i\={u}t\.{e} et al. (2007)  \\
NGC 1817 & 186.16 & $-$13.09 & 0.41 & $-$0.33 &  9.9 & 11.79 & 0.25 &
Parisi et al. (2005)  \\
NGC 2360 & 229.81 & $-$01.43 & 0.56 & $-$0.12 &  9.3 & 11.72 & 0.11 &
Claria et al. (2008)  \\
NGC 2506 & 230.56 & $+$09.93 & 1.11 & $-$0.32 & 10.5 & 12.95 & 0.08 & Henderson
et al. (2007)  \\

\hline
\end{tabular}
\end{table*}

\begin{table*}
 \centering
 \caption{ The observed stars. } 
 \label{tab2}
\begin{tabular}{lccccccl}
  \hline
\multicolumn{1}{l}{Cluster}&  
\multicolumn{1}{c}{Star ID}&  
\multicolumn{1}{c}{$\alpha(2000.0)$}&  
\multicolumn{1}{c}{$\delta(2000.0)$}&  
\multicolumn{1}{c}{V}&  
\multicolumn{1}{c}{B-V}&
\multicolumn{1}{c}{$RV_{helio}$} &
\multicolumn{1}{l}{S/N}\\
\multicolumn{1}{c}{}&
\multicolumn{1}{c}{}&
\multicolumn{1}{c}{(hh mm s)}&
\multicolumn{1}{c}{($\degr$ $\arcsec$ $\arcmin$)}&
\multicolumn{1}{c}{(mag.)}&
\multicolumn{1}{c}{(mag.)}&
\multicolumn{1}{c}{(km s$^{-1}$)} &
\multicolumn{1}{l}{at 6000 \AA } \\
\hline

NGC  752 &  77 & 01 56 21.60 & 37 36 08.00 &  9.35 &$+$1.02 & +6.3$\pm$0.2 & 100 \\
         & 137 & 01 57 03.10 & 38 08 02.00 &  8.89 &$+$1.02 & +5.9$\pm$0.2 & 100 \\
         & 295 & 01 58 29.80 & 37 51 37.00 &  9.29 &$+$0.96 & +6.3$\pm$0.2 & 120 \\
         & 311 & 01 58 52.90 & 37 48 57.00 &  9.04 &$+$1.03 & +6.7$\pm$0.2 & 120 \\
NGC 1817 &1027 & 05 12 19.38 & 16 40 48.64 & 12.13 &$+$1.03 &+66.1$\pm$0.2 & 110 \\
         &2038 & 05 12 06.27 & 16 38 15.34 & 12.17 &$+$1.12 &+66.6$\pm$0.2 &  90 \\
         &2059 & 05 12 24.65 & 16 35 48.84 & 12.04 &$+$1.08 &+66.5$\pm$0.2 &  90 \\
NGC 2360 & 5   & 07 18 14.13 &$-$15 37 30.49 & 10.74 &$+$1.04 &+30.4$\pm$1.1 & 77 \\
         & 6   & 07 18 19.08 &$-$15 37 32.62 & 11.03 &$+$1.04 &+29.1$\pm$0.1 & 95 \\ 
         & 8   & 07 18 10.84 &$-$15 34 13.30 & 11.09 &$+$1.01 &+29.2$\pm$0.2 & 80 \\
         & 12  & 07 18 09.58 &$-$15 31 39.80 & 10.34 &$+$1.16 &+29.5$\pm$0.4 & 75 \\ 
NGC 2506 &2212 & 08 00 08.68 &$-$10 46 37.50 & 11.95 &$+$1.07 &+84.1$\pm$0.2 & 50 \\
         &3231 & 07 59 55.77 &$-$10 48 22.73 & 13.12 &$+$0.98 &+84.9$\pm$0.4 & 45 \\
         &4138 & 08 00 01.49 &$-$10 45 38.50 & 13.30 &$+$0.91 &+84.9$\pm$0.3 & 60 \\
\hline
\end{tabular}
\end{table*}

\section{Observations and data reduction}

Clusters were selected from the \emph{New catalogue of optically visible open
clusters and 
candidates\footnote{http://www.astro.iag.usp.br/~wilton/}} (Dias et al. 2002). 
Emphasis was placed on OCs not yet subjected to high resolution spectroscopy.
Since the main sequence stars in the chosen OCs were faint, we elected to observe
the red 
giant members. 
For each of the target clusters,  red giants were selected using the 
{\small WEBDA\footnote{http://www.univie.ac.at/webda/}} database.
Target clusters and their properties are shown in the Table 1: column 1
represents the cluster name, 
columns 2 \& 3 the Galactic longitude and latitude in degrees, 
column 4 the age, 
column 5 the photometric estimate of the iron abundance, column 6 the Galactocentric
distance, column 7 the
distance modulus, 
column 8 the reddening, column 9 the reference to [Fe/H]. All quantities
are from the database entry except for the [Fe/H] abundance and 
the Galactocentric distance, Rgc, which we calculate assuming 
a distance of the Sun from the Galactic centre of 8.0$\pm$0.6 kpc (Ghez et al. 2008) .   

\begin{figure} 
\begin{center} 
 \vskip-4ex
\includegraphics[width=11.4cm,height=11.5cm]{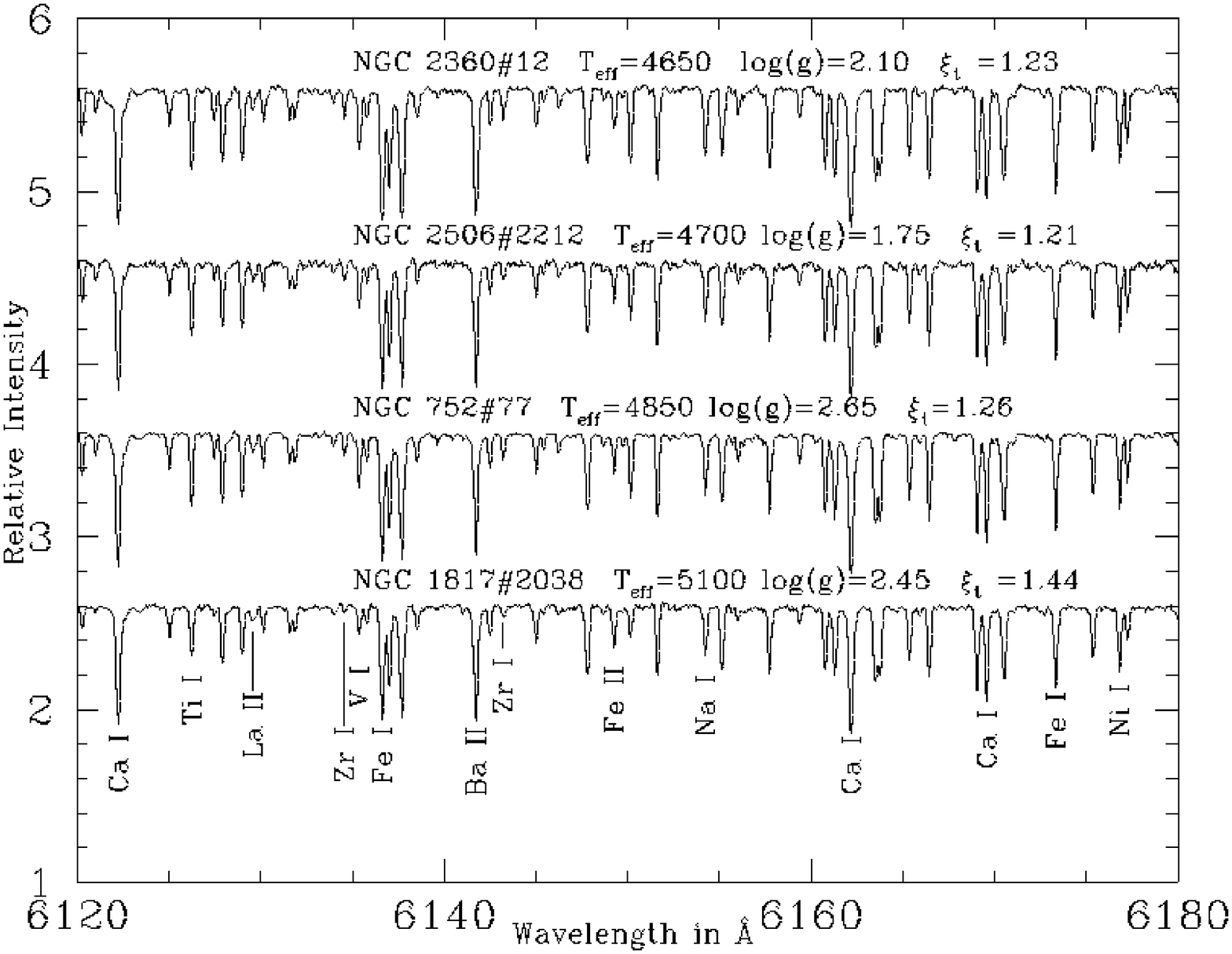} \vskip-11ex
\caption{ Representative spectra for the four clusters.}
 \label{representative} 
 \end{center}
\end{figure}

Observations were conducted during February 6-10, 1999 with Tull echelle coud\'{e}
spectrograph (Tull et al. 1995) on the 2.7-m Harlan J. Smith telescope at the
McDonald observatory.
Spectra are at a resolving power of
$\ga$50,000 as  measured by the FWHM of Th I lines in comparison spectra. 
The spectral coverage was complete from 4000 \AA~ to 5600 \AA~ and substantial but
incomplete from 5600 \AA~ to about 9800 \AA. 
A series of bias and flat frames were obtained at several exposure levels chosen to 
match those of the program stars, and comparison Th-Ar spectra were taken to
establish the wavelength scale. 
We  obtained  on each night
the spectrum of a rapidly-rotating B star to monitor the presence of telluric lines.

Spectroscopic reductions were done with the \footnotesize${\bf IRAF}$ software of 
$NOAO$\footnote{http://iraf.noao.edu/} within 
the \textit{imred} and \textit{echelle} packages, involving bias subtraction,
scattered light correction, 
flat-fielding, 
 wavelength calibration and
continuum fitting.   
We measured the radial velocity (RV) of each star on each spectrum.
The continuum-fitted spectrum was corrected for the Doppler shift using the routine
\textit{dopcor} 
available in \footnotesize{$\textbf{IRAF}$} task 
\textit{splot}. Our radial velocity measurements are in fair agreement with  
the previous radial velocity measurements for the red giants in OCs
(Mermilliod et al. 2008). 

The identification and basic observational data for the stars observed in each of
the clusters are given in Table 2, along with the 
computed radial velocity and S/N of each of the spectra extracted at 6000 \AA\ for each of the stars.
 Spectra of a representative region are shown in Figure 1
for one star from
each of the four clusters.

\section{Abundance analysis}
\subsection{\small Line selection}

Selection of stellar lines which are free from blends is crucial for deriving
accurate elemental abundances. 
We used  Rowland's preliminary table of solar spectrum wavelengths (Moore et al.
1966) and the 
the Arcturus spectrum (Hinkle et al. 2000) to identify  unblended spectral lines.
We employed strict criteria in the selection of suitable lines. First, in order to
avoid the difficulty 
in defining the continuum due to
heavy line crowding in the blue part of the spectrum, we selected lines only within the
4300
to 7850 \AA\ wavelength range. Second, regions containing telluric absorption lines
were generally
avoided.
Third, lines which appear asymmetric  were assumed to be blended with unidentified
lines and
discarded. 
Fourth, lines  with equivalent widths (EWs) below 10 m\AA\ were rejected because
they are too sensitive to noise 
and the normalization of the continuum, and lines with
equivalent widths greater than 230 m\AA\ were discarded because they are too
saturated.

Application of our criteria resulted in a list of 55 Fe I  lines with lower
excitation potentials (LEP) ranging from 0.9 to 5.0 eV and EWs of 
up to 180 m\AA and 10 Fe II lines with excitation potentials of 2.8 to 3.9
eV and EWs up to 120 m\AA.     

A portion of final linelist with solar EWs is given in Table 3 with details about the lines 
including the $\log$ gf's (see below). 
For each line in the constructed linelist, we provide a gf-value from the
literature. In most cases, we located recent experimental determinations or
chose values from critical reviews. References to the adopted sources are given
in Table 3. 
EWs were measured manually using the task '\textit{splot}' contained in
\footnotesize{$\textbf{IRAF}$} 
by fitting a Gaussian profile to the observed line. 
           
For several absorption features, multiple components of a given atomic transition
contribute to the
feature. In such cases, we computed 
a synthetic spectrum including all components and occasionally other lines too  and
matched
the synthetic spectrum to the observed spectrum by varying the abundance of the
element in
question.

As a check on the chosen gf-values, we derived solar abundances using the ATLAS9 model
for $T_{eff}$ = 5777 K, log~$g$=4.44 cgs. A microturbulence of $\xi_t$ = 0.95 km
s$^{-1}$
was found from iron lines. 
Solar
equivalent
widths were measured off the solar integrated disk spectrum (Kurucz et al. 1984).
Abundances are given in Table 4 along with those from the
recent review by Asplund et al. (2009). Our abundances for the majority of elements
are in good agreement with Asplund et al.'s. Small differences in abundances are
inevitable, for example, the line lists are not necessarily identical as
to selected lines and/or gf-values and  the adopted solar models are different; ours is
a classical model but Asplund et al. for many elements use a model representing the
solar granulation. For the purposes of determining the stellar abundances, we adopt 
our solar abundances when computing [X/H] and [X/Fe], i.e., our analysis is essentially
a differential one relative to the Sun.

\begin{table} 
  \begin{minipage}{86mm}
  \caption{Adopted linelist. All
the columns are self-explanatory. }
 \label{tab3}
\begin{tabular}{lllllc}
  \hline
\multicolumn{1}{l}{Atom} &  
\multicolumn{1}{l}{Wavelength} &  
\multicolumn{1}{l}{LEP$^{1}$} &  
\multicolumn{1}{l}{$\log~{gf}$} &
\multicolumn{1}{l}{$W_{\lambda,\odot}$} &
\multicolumn{1}{c}{$Ref^{a}$.} \\
\multicolumn{1}{c}{} &
\multicolumn{1}{c}{( \AA)} &
\multicolumn{1}{c}{(eV)} &
\multicolumn{1}{c}{ }    &
\multicolumn{1}{c}{(m\AA)} &
\multicolumn{1}{c}{ } \\
 \hline

Na I & 4668.567  & 2.100 &  -1.31 &  53.4 & NIST \\  
     & 4982.821  & 2.100 &  -0.96 &  79.0 & NIST \\
     & 5688.210  & 2.100 &  -0.45 & 115.4 & NIST \\
     & 6154.224  & 2.100 &  -1.55 &  36.3 & NIST \\
     & 6160.746  & 2.100 &  -1.25 &  55.9 & NIST \\
Mg I & 5711.090  & 4.340 &  -1.72 & 104.5 & NIST\\
     & 6318.708  & 5.108 &  -1.90 &  42.9 & NIST \\
Al I & 7084.564  & 4.020 &  -0.93 &  21.4 & KUR\\
     & 7835.296  & 4.020 &  -0.65 &  41.2 & KUR\\
     & 7836.119  & 4.020 &  -0.47 &  55.4 & KUR\\ 
Si I & 5665.551  & 4.920 &  -2.04 &  39.4 & NIST\\
     & 5645.606  & 4.930 &  -2.14 &  34.9 & LUCK\\
     & 5701.100  & 4.930 &  -2.05 &  37.7 & NIST\\
     & 5753.632  & 5.610 &  -1.30 &  42.5 & LUCK\\
     & 6131.570  & 5.610 &  -1.70 &  22.3 & LUCK \\
     & 6131.851  & 5.610 &  -1.62 &  23.3 & LUCK\\
     & 6145.013  & 5.610 &  -1.48 &  36.5 & LUCK\\
     & 6237.319  & 5.610 &  -1.14 &  56.7 & LUCK\\
     & 6244.470  & 5.610 &  -1.36 &  45.1 & LUCK\\
     & 6243.812  & 5.613 &  -1.26 &  46.5 & KUR\\
     & 6142.486  & 5.620 &  -1.54 &  33.0 & LUCK\\
     & 6721.840  & 5.862 &  -1.06 &  42.8 & LUCK\\
     & 6195.445  & 5.873 &  -1.80 &  14.9 & LUCK\\
Ca I & 6122.221  & 1.890 &  -0.32 & 161.9 & LUCK\\
     & 5581.971  & 2.520 &  -0.55 &  93.7 & LUCK\\
     & 5590.117  & 2.520 &  -0.57 &  91.7 & LUCK\\
     & 6166.434  & 2.520 &  -1.14 &  69.2 & LUCK\\
     & 6169.560  & 2.520 &  -0.48 & 108.5 & LUCK\\
     & 6455.599  & 2.520 &  -1.29 &  56.2 & LUCK\\
     & 6499.649  & 2.520 &  -0.82 &  85.0 & LUCK\\
     & 6471.662  & 2.526 &  -0.68 &  90.7 & LUCK\\
Sc I & 5686.832  & 1.440 &   0.38 &   8.3 & LUCK\\
     & 5356.090  & 1.860 &   0.17 &   2.3 & LUCK\\
Sc II& 6604.587  & 1.357 &  -1.31 &  35.1 & NIST\\
     & 5667.141  & 1.500 &  -1.20 &  30.0 & KUR \\
     & 6245.615  & 1.507 &  -1.03 &  35.4 & KUR\\   
     & 6300.681  & 1.507 &  -1.89 &   8.1 & KUR \\  
     & 5526.813  & 1.768 &   0.06 &  75.6 & KUR\\
Ti I & 5039.960  & 0.021 &  -1.13 &  75.7 & NIST\\
     & 5460.497  & 0.048 &  -2.75 &   9.6 & LUCK \\
     & 4999.510  & 0.826 &   0.31 & 103.6 & LUCK\\
     & 5020.026  & 0.836 &  -0.35 &  72.6 & KUR\\
     & 5295.776  & 1.067 &  -1.63 &  13.1 & NIST\\
     & 5474.223  & 1.460 &  -1.23 &  10.8 & NIST\\
     & 5490.148  & 1.460 &  -0.93 &  21.6 & NIST \\
     & 4617.274  & 1.749 &   0.39 &  61.2 & NIST\\
     & 5739.980  & 2.236 &  -0.60 &   7.3 & NIST\\
     & 5702.656  & 2.292 &  -0.57 &   8.1 &  NIST\\\
Ti II& 4764.528  & 1.237 &  -2.77 &  37.2 & LUCK\\  
     & 4708.665  & 1.240 &  -2.21 &  52.9 & NIST\\  
     & 5005.168  & 1.566 &  -2.54 &  25.5 & NIST\\ 
     & 5381.022  & 1.566 &  -1.85 &  57.3 & LUCK\\ 
     & 5396.244  & 1.580 &  -2.92 &  12.1 & LUCK\\  
     & 5336.788  & 1.582 &  -1.70 &  69.2 & NIST \\
     & 5418.767  & 1.582 &  -1.99 &  48.1 & KUR \\ 
V I  & 6251.823  & 0.286 &  -1.34 &  15.8 & NIST\\
     & 6111.647  & 1.043 &  -0.71 &  10.7 & NIST\\
     & 5727.653  & 1.051 &  -0.87 &   8.9 & NIST\\
     & 6135.366  & 1.051 &  -0.75 &  10.4 & NIST\\
     & 5737.062  & 1.064 &  -0.74 &  10.8 & NIST\\
     & 5668.365  & 1.081 &  -1.03 &   5.6 & NIST\\
     & 5670.848  & 1.081 &  -0.42 &  19.0 & NIST\\
     & 5727.044  & 1.081 &  -0.01 &  39.9 & NIST\\

\end{tabular}
\end{minipage} 
\end{table}

\begin{minipage}{90mm}
\flushleft{{\bf Table 3}$-$ continued}
\begin{tabular}{lllllc}
  \hline
\multicolumn{1}{l}{Atom} &  
\multicolumn{1}{l}{Wavelength} &  
\multicolumn{1}{l}{LEP$^{1}$} &  
\multicolumn{1}{l}{$\log~{gf}$} &
\multicolumn{1}{l}{$W_{\lambda,\odot}$} & \multicolumn{1}{c}{$Ref^{a}$.} \\ 
\multicolumn{1}{c}{} &
\multicolumn{1}{c}{(\AA)} &
\multicolumn{1}{c}{(eV)} &
\multicolumn{1}{c}{ }    &
\multicolumn{1}{c}{(m\AA)} & \multicolumn{1}{c}{ }\\
 \hline
 
Cr I & 4545.958  & 0.941 &  -1.37 &  81.1 & SLS\\
     & 5296.696  & 0.983 &  -1.36 &  91.0 & SLS\\
     & 5300.747  & 0.983 &  -2.00 &  58.3 & SLS\\
     & 5345.802  & 1.004 &  -0.95 & 112.2 & SLS\\
     & 5238.959  & 2.709 &  -1.30 &  16.1 & SLS\\
     & 5329.139  & 2.910 &  -0.06 &  65.8 & KUR\\
     & 5784.967  & 3.321 &  -0.38 &  31.2 & NIST\\
     & 5214.129  & 3.369 &  -0.74 &  17.1 & NIST\\
     & 5628.640  & 3.422 &  -0.74 &  14.4 & SLS\\
     & 5287.174  & 3.440 &  -0.87 &  10.7 & SLS\\   
     & 5312.853  & 3.449 &  -0.55 &  19.8 & SLS\\
     & 5304.178  & 3.463 &  -0.67 &  15.4 & SLS\\
Cr II& 5279.874  & 4.070 &  -2.10 &  19.0 & NIST\\ 
     & 5308.424  & 4.071 &  -1.81 &  25.3 & NIST\\
     & 5237.321  & 4.073 &  -1.16 &  52.6 & NIST\\
     & 5334.864  & 4.073 &  -1.56 &  33.6 & KUR\\
     & 5313.578  & 4.074 &  -1.65 &  33.4 & NIST\\ 
     & 5502.081  & 4.170 &  -1.99 &  18.3 & NIST\\
Mn I & 6013.488  & 3.072 &  -0.25 &  84.9 & NIST\\
     & 6021.796  & 3.075 &   0.03 &  90.7 & NIST\\
     & 5377.608  & 3.845 &  -0.11 &  48.0 & KUR\\
     & 5399.480  & 3.850 &  -0.29 &  37.4 & KUR\\
Fe I & 6136.995  & 2.198 &  -2.95 &  65.6 & F\&W\\
     & 6252.562  & 2.404 &  -1.69 & 120.9 & F\&W \\
     & 5141.742  & 2.424 &  -2.24 &  84.5 & F\&W\\
     & 5701.548  & 2.559 &  -2.22 &  83.8 & F\&W \\
     & 6646.931  & 2.608 &  -3.95 &   9.3 & KUR\\
     & 5036.918  & 3.018 &  -3.04 &  24.5 & F\&W\\
     & 5215.184  & 3.266 &  -0.87 & 119.7 & F\&W\\
     & 5576.093  & 3.431 &  -0.94 & 106.2 & F\&W\\
     & 5568.863  & 3.635 &  -2.95 &  10.3 & LUCK\\
     & 5054.655  & 3.640 &  -1.92 &  39.8 & F\&W \\
     & 5636.695  & 3.640 &  -2.56 &  20.7 & F\&W\\
     & 5760.343  & 3.642 &  -2.44 &  22.9 & F\&W\\
     & 5539.278  & 3.643 &  -2.61 &  18.7 & F\&W \\
     & 6411.653  & 3.654 &  -0.72 & 116.9 & F\&W\\
     & 5466.986  & 3.655 &  -2.23 &  32.5 & F\&W\\ 
     & 6336.828  & 3.687 &  -0.86 & 101.9 & F\&W\\
     & 5379.574  & 3.695 &  -1.51 &  60.5 & F\&W\\
     & 6003.014  & 3.882 &  -1.15 &  81.6 & NIST\\
     & 6187.988  & 3.943 &  -1.67 &  46.0 & F\&W \\
     & 5293.957  & 4.143 &  -1.84 &  28.7 & F\&W \\
     & 6165.358  & 4.143 &  -1.47 &  43.9 & F\&W \\
     & 5608.973  & 4.209 &  -2.40 &  10.4 & LUCK\\
     & 5618.631  & 4.209 &  -1.28 &  49.6 & F\&W\\
     & 5074.753  & 4.221 &  -0.23 & 114.3 & F\&W\\    
     & 5738.230  & 4.221 &  -2.34 &  11.8 & LUCK\\
     & 5579.338  & 4.231 &  -2.40 &  10.6 & LUCK\\
     & 5016.477  & 4.256 &  -1.69 &  32.7 & LUCK\\
     & 5090.783  & 4.256 &  -0.44 &  90.1 & NIST\\     
     & 5243.777  & 4.256 &  -1.12 &  59.9 & F\&W\\
     & 5646.682  & 4.261 &  -2.50 &   7.1 & LUCK\\
     & 5717.832  & 4.284 &  -1.10 &  60.5 & F\&W\\
     & 5197.934  & 4.301 &  -1.62 &  35.5 & F\&W\\
     & 5466.398  & 4.371 &  -0.63 &  77.0 & LUCK\\ 
     & 5295.312  & 4.415 &  -1.67 &  28.5 & F\&W \\
     & 5560.210  & 4.435 &  -1.16 &  50.5 & F\&W\\
     & 5577.022  & 5.033 &  -1.55 &  12.3 & LUCK\\
Fe II& 5000.730  & 2.780 &  -4.61 &  11.9 & M\&B \\   
     & 4520.225  & 2.807 &  -2.65 &  81.3 & M\&B\\
     & 4993.353  & 2.807 &  -3.62 &  39.0 & M\&B\\  
     & 6369.460  & 2.891 &  -4.11 &  19.5 & M\&B \\  
     & 6432.680  & 2.891 &  -3.57 &  40.9 & M\&B\\ 
\hline
\end{tabular}
\end{minipage}

 \begin{minipage}{90mm}
 \flushleft{{\bf Table 3}$-$ continued}
 \begin{tabular}{llllcc}
  \hline
\multicolumn{1}{l}{Atom} &  
\multicolumn{1}{l}{Wavelength} &  
\multicolumn{1}{l}{LEP$^{1}$} &  
\multicolumn{1}{l}{$\log~{gf}$} &
\multicolumn{1}{l}{$W_{\lambda,\odot}$} & \multicolumn{1}{c}{$Ref^{a}$.} \\
\multicolumn{1}{c}{} &
\multicolumn{1}{c}{(\AA)} &
\multicolumn{1}{c}{(eV)} &
\multicolumn{1}{c}{ }    &
\multicolumn{1}{c}{(m\AA)} &  \multicolumn{1}{c}{ }\\
\hline

     & 5256.931  & 2.892 &  -4.06 &  21.1 & M\&B\\
     & 5425.260  & 3.199 &  -3.22 &  42.6 & M\&B\\          
     & 6084.103  & 3.199 &  -3.88 &  20.2 & F\&W \\
     & 5234.620  & 3.221 &  -2.18 &  84.8 & M\&B\\          
     & 5414.067  & 3.221 &  -3.58 &  27.7 & M\&B\\
     & 6149.244  & 3.889 &  -2.84 &  35.8 & F\&W\\
     & 6247.559  & 3.892 &  -2.43 &  52.1 & F\&W\\
     & 6456.383  & 3.903 &  -2.19 &  61.0 & F\&W\\
Co I & 6116.994  & 1.785 &  -2.49 &   6.0 & NIST\\
     & 5647.232  & 2.280 &  -1.56 &  13.8 & NIST\\
     & 5212.680  & 3.514 &  -0.14 &  19.2 & KUR\\
     & 6454.990  & 3.632 &  -0.25 &  14.6 & NIST\\
     & 5342.701  & 4.022 &   0.69 &  30.8 & KUR \\
Ni I & 5578.720  & 1.676 &  -2.64 &  55.3 & NIST\\ 
     & 5748.352  & 1.676 &  -3.26 &  27.6 & NIST\\
     & 6191.181  & 1.677 &  -2.35 &  71.0 & KUR\\
     & 6177.242  & 1.826 &  -3.51 &  14.4 & NIST\\
     & 4998.229  & 3.606 &  -0.78 &  52.9 & NIST\\
     & 5462.493  & 3.847 &  -0.93 &  38.6 & NIST\\
     & 5468.103  & 3.847 &  -1.61 &  12.4 & NIST\\
     & 5589.357  & 3.898 &  -1.14 &  27.6 & NIST\\
     & 5593.736  & 3.898 &  -0.84 &  40.2 & NIST\\
     & 5638.745  & 3.898 &  -1.73 &   9.5 & NIST\\
     & 6111.072  & 4.088 &  -0.87 &  32.7 & NIST\\
     & 5625.316  & 4.089 &  -0.70 &  37.7 & NIST \\
     & 5682.199  & 4.105 &  -0.47 &  50.4 & NIST\\
     & 5760.830  & 4.105 &  -0.80 &  33.9 & NIST\\
Cu I & 5218.201  & 3.820 &   0.26 &  49.3 & NIST\\
Zn I & 4722.160  & 4.030 &  -0.34 &  65.0 & KUR\\
     & 6362.342  & 5.790 &   0.27 &  28.1 & LUCK\\
Y II & 5200.409  & 0.992 &  -0.57 &  36.3 & HLG\\ 
     & 4982.133  & 1.033 &  -1.29 &  13.3 & HLG\\
     & 5289.817  & 1.033 &  -1.85 &   4.0 & HLG\\
     & 4883.688  & 1.080 &   0.07 &  54.3 & HLG\\
     & 5402.773  & 1.840 &  -0.63 &  12.3 & LUCK\\ 
Zr I & 6143.201  & 0.070 &  -1.10 &   2.1 & BGHL\\
     & 4739.483  & 0.650 &   0.23 &   6.2 & BGHL\\
Ba II& 5853.678  & 0.604 &  -1.02 &  62.5 & LUCK\\
     & 6496.905  & 0.604 &  -0.37 &  97.8 & LUCK\\
La II& 5303.538  & 0.321 &  -1.35 &   4.4 & LBS\\
     & 6390.486  & 0.321 &  -1.41 &   3.0 & LBS\\
Ce II& 5472.281  & 1.240 &  -0.18 &   2.1 & LUCK\\    
Nd II& 5092.800  & 0.380 &  -0.61 &   6.5 & DLS\\
     & 5319.813  & 0.550 &  -0.14 &  11.4 & DLS\\
     & 5485.539  & 1.260 &  -0.12 &   4.1 & DLS\\
Sm II& 4519.630  & 0.544 &  -0.35 &   6.1 & LDS\\
Eu II& 6645.108  & 1.379 &   0.12 &   4.80 & LWD \\

\hline
\end{tabular}
\flushleft{$^{1}$ The lines are arranged in the order of their increasing LEP.} 
 
\begin{flushleft}
             $^{a}$References for the adopted gf-values: \\
               BGHL$-$Bi\'{e}mont et al. (1981);\\
               F\&W$-$Fuhr \& Wiese (2006);\\
               HLG$-$Hannaford et al. (1982);\\
               KUR$-$Kurucz (1998); \\
               LWD$-$Lawler et al. (2001); \\
               LBS$-$Lawler et al. (2001); \\
               LDS$-$Lawler et al. (2006); \\
               LSC$-$Lawler et al. (2009);\\
               LUCK$-$Luck (Private communication); \\
               M\&B$-$Mel\'{e}ndez \& Barbuy (2009); \\
               SLS$-$Sobeck et al. (2007);\\ 
               NIST$-$Atomic Spectra Database
\footnote{http://physics.nist.gov/PhysRefData/ASD/lines$\_$form.html } 
\end{flushleft}
\end{minipage}

\begin{table}
 \centering
 \begin{minipage}{85mm}
  \caption{Solar abundances derived by employing the solar model atmosphere from
Castelli \& Kurucz (2003) compared with  
           the photospheric abundances from 
Asplund et al. (2009). }
 \label{tab4}
\begin{tabular}{ccc}
  \hline
\multicolumn{1}{c}{Species}&  
\multicolumn{1}{c}{$\log~\epsilon_{\odot}$}&  
\multicolumn{1}{c}{$\log~\epsilon_{\odot}$} \\
\multicolumn{1}{c}{}&
\multicolumn{1}{c}{(our study) }&
\multicolumn{1}{c}{(Asplund)} \\
 \hline

Na I& 6.29$\pm$0.03(5) &  6.24$\pm$0.04  \\
Mg I& 7.55$\pm$0.06(2) &  7.60$\pm$0.04  \\
Al I& 6.35$\pm$0.06(3) &  6.45$\pm$0.03  \\
Si I& 7.55$\pm$0.05(2) &  7.51$\pm$0.03  \\
Ca I& 6.28$\pm$0.05(8) &  6.34$\pm$0.04  \\
Sc I& 2.98$\pm$0.03(2) &  3.15$\pm$0.04  \\
Sc II& 3.17$\pm$0.05(5) &                \\ 
Ti I&  4.88$\pm$0.06(10)& 4.95$\pm$0.05  \\
Ti II& 4.92$\pm$0.09(7) &                \\
V I &  3.94$\pm$0.05(8) & 3.93$\pm$0.08  \\
Cr I&  5.59$\pm$0.04(12) & 5.64$\pm$0.04  \\ 
Cr II& 5.67$\pm$0.05(6) &                 \\ 
Mn I&  \bf 5.39         &  5.43$\pm$0.04  \\
Fe  I& 7.54$\pm$0.05(36)&  7.50$\pm$0.04  \\
Fe II& 7.52$\pm$0.05(13)&                \\ 
Co I&  4.86$\pm$0.03(5) &  4.99$\pm$0.07  \\ 
Ni I&  6.24$\pm$0.02(14) & 6.22$\pm$0.04  \\ 
Cu I&  \bf 4.18         &  4.19$\pm$0.04  \\
Zn I&  4.59$\pm$0.00(2) &  4.56$\pm$0.05  \\
Y II&  2.19$\pm$0.07(5) &  2.21$\pm$0.05  \\
Zr I&  2.59$\pm$0.11(2) &  2.58$\pm$0.04  \\
Ba II& \bf 2.13         &  2.18$\pm$0.09  \\
La II& 1.24$\pm$0.13(2) &  1.10$\pm$0.04  \\
Ce II& \bf1.56          &  1.58$\pm$0.04  \\
Nd II& 1.50$\pm$0.09(3) &  1.42$\pm$0.04  \\
Sm II& \bf1.00          &  0.96$\pm$0.04  \\
Eu II& \bf 0.51         &  0.52$\pm$0.04  \\

\hline
\end{tabular}
\flushleft{Note: Numbers in the parentheses indicate the number of lines used for
abundance analysis. The abundances calculated by synthesis are presented in
bold numbers.}
\end{minipage}
\end{table}

\subsection{Determination of atmospheric parameters}
\subsubsection{Photometry}

Initial estimates of the effective temperature of a red giant were derived  from
dereddened B and V photometry 
using the empirically-calibrated colour-temperature relation 
by Alonso et al. (1999) based on a large sample of field and globular cluster
giants of spectral types from F0 to K5. An error 
of 3\% is expected in the derived temperatures.

Gravities were computed using the known distance to the OCs, temperature
and bolometric corrections, 
and the cluster turn-off mass.
We have adopted a turn-off mass of 1.5M$_{\odot}$ for NGC 752
(Bartasiute et al. 2007), 
2M$_{\odot}$ for NGC 1817 (Jacobson et al. 2009), 1.98M$_{\odot}$ 
for NGC 2360 (Hamdani et al. 2000), and 1.69M$_{\odot}$ for NGC 2506 (Carretta et
al. 2004).
The relation between log~$g$ and T$_{eff}$ is given by (Allende Prieto et al. 1999) \\

 $\log{g_\star}$= $\log{g}_{\odot}$+$\log(M_{\star}/M_{\odot})$+4 $\log(T_{eff}/T_{eff}{\odot})$
 \begin{equation}
    +0.4(V_{0}+BC_{V})+2 log{\pi}+0.12  
\end{equation}
 with the corresponding luminosity given by
\begin{equation}
 log(L_{*}/L_{\odot})= -[0.4 (V_{0}+BC_{V})+2 log~{\pi}+0.12]  
 \end{equation}   \vskip1ex \noindent
 where $\pi$ is the parallax, V$_{0}$ is the apparent Johnson V magnitude corrected
for reddening, and BC$_{V}$ is the bolometric correction.
We adopt log~{g}$_{\odot}$= 4.44, T$_{eff},_{\odot}$= 5777 K. 
Bolometric corrections were taken from the calibration by Alonso et al. (1999).

We suppose that the errors in different quantities involved in equation (1) are
independent of each other. 
Then, by assuming an error of 10\% in the stellar mass, 
an uncertainty of 3\% in T$_{eff}$, an uncertainty of 5\% in photometric V magnitude
and the bolometric corrections, and an error of 10\% in parallax, we get an
error of $\simeq$ 0.11 dex in $\log~g$ and the corresponding uncertainty in
$\log~L_{*}$ amounts to 0.08.

\subsubsection{Spectroscopy}
The spectroscopic abundance analysis was performed with the 2010 version of the local 
thermodynamical equilibrium (LTE) line synthesis program MOOG (Sneden
1973)\footnote{http://www.as.utexas.edu/~chris/moog.html}.
Model atmospheres were generated by linear interpolation from the ATLAS9 model
atmosphere 
grid (Castelli \& Kurucz 2003)\footnote{http://kurucz.harvard.edu/grids.html}. 
A model atmosphere is 
characterized by the effective temperature T$_{eff}$, the surface gravity $\log~{g}$, 
microturbulence velocity $\xi_{t}$, and composition. 
These models use the classical assumptions of line-blanketed plane-parallel 
uniform atmospheres in LTE and hydrostatic equilibrium with flux conservation.   

\begin{figure*} 
\begin{center} 
\includegraphics[width=24cm,height=18cm]{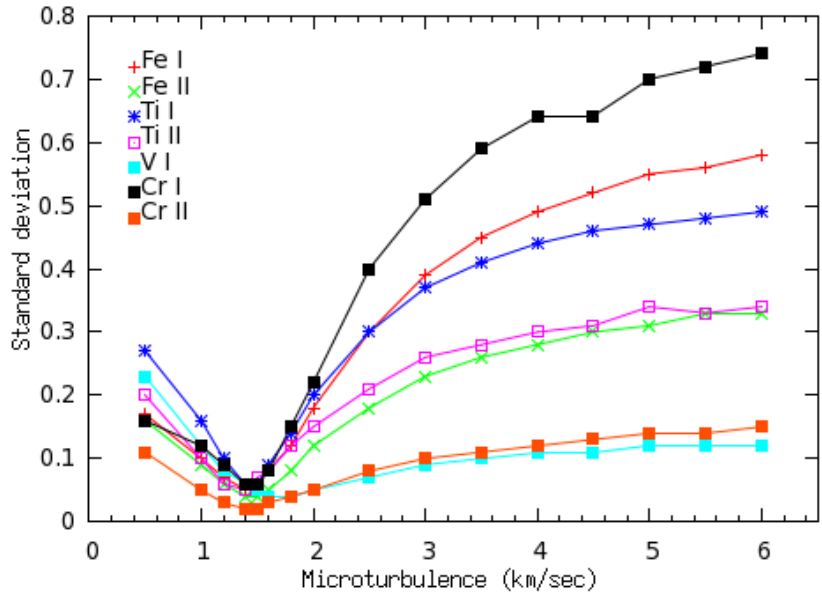} \vskip-62ex 
\caption{The standard deviation for various species about the mean abundances as a
function 
of microturbulence for the star NGC 752 \#311. }
  \label{micro_turb}  
 \end{center}
\end{figure*}

\begin{figure*} 
\begin{center} 
\includegraphics[width=20.5cm,height=14.5cm]{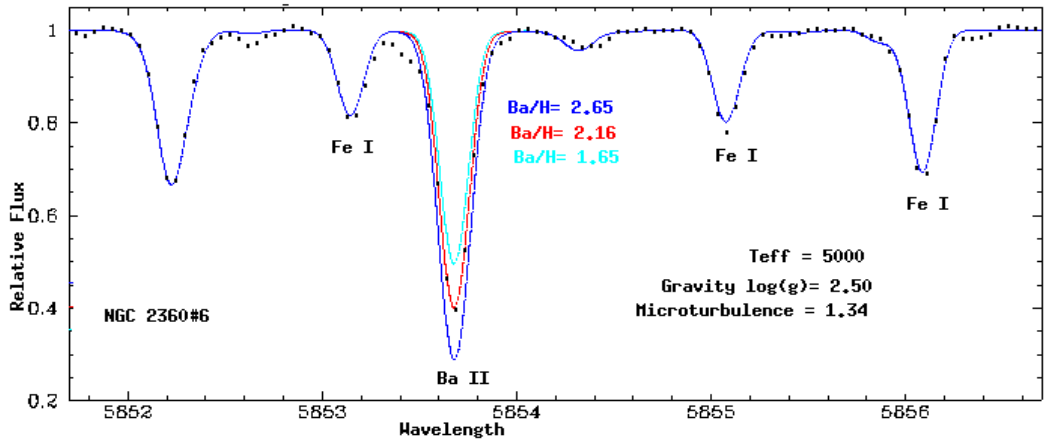} \vskip-56ex 
\caption{Synthetic spectra and the observed spectrum of NGC 2360 \#6 around the
Ba II line at 5853 \AA. The indicated abundances in the figure are on a logarithmic scale.}
  \label{synthba}  
 \end{center}
\end{figure*}

\begin{figure*} 
\begin{center}  \vskip-7ex
\includegraphics[width=20.5cm,height=15.5cm]{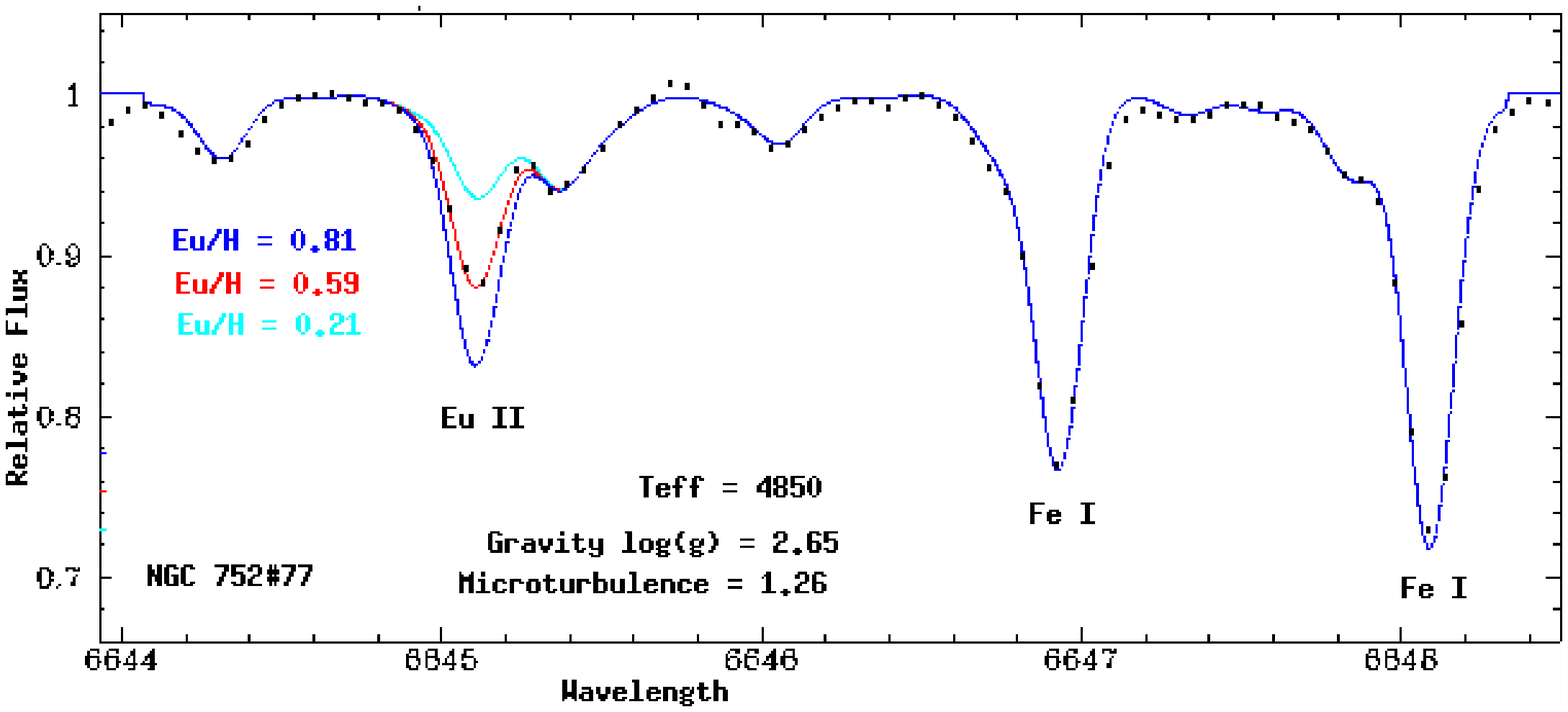} \vskip-65ex 
\caption{Synthetic spectra and the observed spectrum of NGC 752 \#77 around the Eu II
line at 6645 \AA. The indicated abundances in the figure are on a logarithmic scale.}
  \label{syntheu}  
\end{center} 
\end{figure*}

Spectroscopically, we determine the stellar parameters in the conventional way when LTE
is the paramount assumption. The key lines are those of Fe I and Fe II
for which we take gf-values from Fuhr \& Wiese (2006) and Mel\'{e}ndez \& Barbuy
(2009).
The microturbulence assumed to be isotropic and depth 
independent is determined from Fe II lines by the requirement that the
abundance be independent of a line's EW. A model atmosphere with the photometrically
determined parameters was used initially for this determination. 
The effective temperature is found by imposing the requirement that the Fe abundance
from
Fe I lines be independent of a line's lower excitation potential. Finally, the
surface gravity is found by requiring that Fe I and Fe II lines give
the
same Fe abundance for the derived effective temperature and microturbulence.  

A check on the microturbulence is provided by lines of species other than
Fe I. For example, for the star NGC 752 \#311 we 
show  in Figure \ref{micro_turb} the dispersion of the abundance computed
from the Fe I, Fe II, Ti I, Ti II, V I, Cr I and Cr II lines as the micorturbulence
is varied 
over the range in the microturbulence, $\xi_{t}$, from 0 to 6 km s$^{-1}$.
It is clear that the minimum value of dispersion for all species is in the range
1.2-1.6 km s$^{-1}$.
Thus, we adopt a microturbulence of 1.45 km s$^{-1}$ with an uncertainty of 0.20
km s$^{-1}$.

 Several elements other than Fe provide both neutral and
ionized
lines and so offer a check on the condition of ionization equilibrium  of Fe.
Consider for
example the four giants from NGC 752: the abundance differences [X/H] between neutral
and ionized lines of Sc, Ti, V  and Cr are on average 0.03, -0.03, -0.01, and -0.05
dex,
respectively where $\pm$0.05 dex corresponds to a change of $log g$ by $\mp$ of 0.15.

The uncertainties in the derived surface temperatures from spectroscopy are provided by the
errors in the slope of the relation between the Fe I abundance and LEP of the lines. A 
perceptible change of slope occures for variations of the temperature from 50$-$100 K about 
the adopted model. 

Therefore, the typical errors considered in this analysis are 100 K in $T_{eff}$, 0.25 cm s$^{-2}$ 
in log~$g$ and 0.20 km s$^{-1}$ in $\xi_{t}$.

The derived stellar parameters for program stars in each of the cluster are given in 
Table \ref{tab5}: column 1 represents the cluster name, column 2 the star ID, 
columns 3 \& 4 the photometric T$_{eff}$ and log~$g$ values, 
columns 5-7 the spectroscopic T$_{eff}$, log~$g$ and $\xi_{t}$ estimates. 
Finally, the spectroscopic and 
photometric luminosities ( log(L/L$_{\odot})$) are presented in columns 8 \& 9.  
Photometric and spectroscopic estimates are in excellent agreement. 
Mean differences in T$_{eff}$, log~$g$ and logL/L$_{\odot}$ across the 14 stars are
$-28\pm105$ K, $0.06\pm0.10$ cgs, and $-0.08\pm0.10$, respectively.  
The corresponding comparison of the spectroscopic with the photometric [Fe/H] in
Table 1 also illustrates 
fair agreement: $\Delta$[Fe/H] = 0.09 (NGC 752), 0.21 (NGC 1817), 0.05 (NGC 2360), and
0.12 (NGC 2506).

\begin{table*}
\centering
 \begin{minipage}{150mm}
 \caption{Basic photometric and spectroscopic atmospheric parameters for the stars
in each cluster.}
 \label{tab5}
 \begin{tabular}{lcccccccc}
  \hline
\multicolumn{1}{l}{Cluster}&  
\multicolumn{1}{c}{Star ID}&  
\multicolumn{1}{c}{$T_{eff,phot}$}&  
\multicolumn{1}{c}{$\log g_{phot}$}&  
\multicolumn{1}{c}{$T_{eff,spec}$}&  
\multicolumn{1}{c}{$\log g_{spec}$}&
\multicolumn{1}{c}{$\xi_{spec}$}&
\multicolumn{1}{c}{$\log(L/L_\odot)$} &
\multicolumn{1}{l}{$\log(L/L_\odot)$} \\
\multicolumn{1}{c}{}&
\multicolumn{1}{c}{}&
\multicolumn{1}{c}{(K)}&
\multicolumn{1}{c}{(cm s$^{-2}$)}&
\multicolumn{1}{c}{(K)}&
\multicolumn{1}{c}{(cm s$^{-2}$)}&
\multicolumn{1}{c}{(km sec$^{-1}$)}& \multicolumn{1}{c}{spectroscopy} &
\multicolumn{1}{l}{photometry} \\

\hline

NGC  752 &  77 & 4780 & 2.75  & 4850 & 2.65 & 1.26 & 1.71 & 1.54 \\
         & 137 & 4780 & 2.57  & 4850 & 2.50 & 1.36 & 1.81 & 1.72 \\
         & 295 & 4899 & 2.80  & 5050 & 2.85 & 1.47 & 1.53 & 1.53 \\
         & 311 & 4761 & 2.62  & 4850 & 2.60 & 1.45 & 1.71 & 1.66 \\
NGC 1817 &1027 & 5177 & 2.66  & 5100 & 2.60 & 1.39 & 1.92 & 1.89 \\
         &2038 & 4968 & 2.57  & 5100 & 2.45 & 1.44 & 2.07 & 1.90 \\
         &2059 & 5059 & 2.57  & 4800 & 2.40 & 1.38 & 2.01 & 1.94 \\
NGC 2360 & 5   & 4899 & 2.56  & 4900 & 2.70 & 1.29 & 1.75 & 1.89 \\
         & 6   & 4899 & 2.68  & 5000 & 2.50 & 1.34 & 1.98 & 1.77 \\ 
         & 8   & 4962 & 2.74  & 5050 & 2.60 & 1.37 & 1.90 & 1.74 \\
         & 12  & 4668 & 2.27  & 4650 & 2.10 & 1.23 & 2.26 & 2.10 \\
NGC 2506 &2212 & 4710 & 1.86  & 4700 & 1.75 & 1.21 & 2.56 & 2.45 \\
         &3231 & 4893 & 2.44  & 5000 & 2.50 & 1.42 & 1.92 & 1.94 \\
         &4138 & 5048 & 2.60  & 5100 & 2.60 & 1.47 & 1.85 & 1.84 \\

\hline
 \end{tabular}
 \end{minipage}
\end{table*}

\subsection{Synthetic spectra}
We compared the stellar spectra to the synthetic spectra to derive abundances for
the lines 
having intrinsic multiple components and  lines affected 
by blends.
Figures \ref{synthba} and \ref{syntheu} show synthetic spectra fit to the observed one 
using three different abundances. The dotted line is the stellar spectrum. The red
line is the best 
fit to the stellar spectrum, 
with the other lines representing different values for [Ba/H] and [Eu/H] 
abundances, based on $\chi^{2}$ goodness of fit provided by MOOG.      

In this analysis, we have adopted the hfs data of Prochaska \& McWilliam (2000) for 
the synthesis of Mn I line at 6013 \AA, Allen et al. (2011) for Cu I line at 5218 \AA,
McWilliam (1998) for Ba II line at 5853 \AA\ and Mucciarelli et al. (2008) for Eu II
line 6645 \AA. Isotopic ratios for Cu I, Ba II and Eu II were taken from Lodders (2003).
Further, we have synthesized the lines Ce II line at 5472 \AA\ and Sm II line at
4577 \AA\ since the 
blends make it impossible to measure their EWs.
The spectrum synthesis was carried out by running the MOOG in '\textit{synth}' mode.
 
Since all odd species exhibit hfs effects of relatively varying strengths, we have performed spectrum
synthesis over Sc II line at 6245 \AA\, V I line at 5727 \AA\ and Co I line at 5647 \AA\ .
Here, we have adopted the hfs data of Prochaska \& McWilliam (2000) for Sc II and for V I and Co I hfs
components were taken from Kurucz linelists\footnote{http://kurucz.harvard.edu/linelists.html}.
We noticed that the these lines are not severely effected by hfs effects, causing an 
abundance difference of 0.0$-$0.10 dex with and 
without the inclusion of hfs components,  
and negligible while considering the standard deviation around mean what we obtain in 
the fine analysis using the routine 'abfind' in MOOG.

\subsection{Abundances}

The abundance analysis was conducted with the model atmospheres having the stellar
parameters determined from the spectra (Table 2),  the line list (Table 12) 
and the program MOOG. Abundances [X/H] are expressed relative to the solar
abundances derived from the adopted gf-values. Results for the individual stars
in each of the OCs are given in Tables 8, 9, 10, and 11.

For each abundance based on analysis of EWs, the abundance and the standard deviation
 were calculated from all lines of a given species. 
 The tables give the abundances of [Fe/H] and [X/Fe] 
for all elements. The quantity [X/Fe] minimizes the sensitivity to errors
in the model atmosphere arising from uncertainties affecting the stellar
parameters.  Inspection of the Tables 8-11 shows that, in general, the
compositions [X/Fe] of stars in a given cluster are  generally identical to within the
(similar) standard deviations computed for an individual star.  Exceptions tend to
occur for species represented by few lines, as expected when the uncertainty in
measuring equivalent widths is a contributor to the total uncertainty.  
From the spread in the abundances for the stars of a given cluster we obtain
the standard deviation $\sigma_1$ in the Tables 8-11 in the column headed
`average'. 
Errors in the adopted gf values are unimportant when providing differential
abundances ([X/H] or [X/Fe]) provided that the solar and stellar abundances depend
on the same set of lines.

\begin{table}
\centering
 \begin{minipage}{90mm}
 \caption{Sensitivity of abundances to the uncertainties in the model
 parameters for the star 5 in NGC 2360 with $T_{eff}$= 4900 K, $\log{g}$= 2.70
cm s$^{-2}$,and $\xi_{t}$= 1.29 km s$^{-1}$.}
 \label{sensitivity}
 \begin{tabular}{lllll}
  \hline
\multicolumn{1}{l}{ }&\multicolumn{1}{l}{$T_{eff}\pm$100
K}&\multicolumn{1}{l}{$\log~{g}\pm$ 0.25} &\multicolumn{1}{l}{$\xi_{t}\pm$ 0.20} &
\multicolumn{1}{l}{ } \\ \cline{2-5}
\multicolumn{1}{l}{Species}&\multicolumn{1}{c}{$\sigma_{T_{eff}}$}&\multicolumn{1}{c}{$\sigma_{log{g}}$}
 & \multicolumn{1}{c}{$\sigma_{\xi_{t}}$} &\multicolumn{1}{c}{$\sigma_{2}$}  \\
\hline

Na I& $+0.06/-0.07$ &$-0.01/+0.01$  &$-0.06/+0.06$   & 0.05  \\
Mg I& $+0.05/-0.05$ &$-0.05/+0.05$  &$-0.06/+0.05$   & 0.05  \\
Al I& $+0.03/-0.04$ &$-0.02/ 0.00$  &$-0.03/+0.03$   & 0.03  \\
Si I& $-0.03/+0.04$ &$+0.04/-0.06$  &$-0.02/+0.03$   & 0.04  \\
Ca I& $+0.08/-0.08$ &$-0.01/ 0.00$  &$-0.11/+0.11$   & 0.08  \\
Sc I& $+0.14/-0.15$ &$+0.01/-0.01$  &$-0.02/+0.03$   & 0.08  \\
Sc II&$-0.03/+0.04$ &$+0.11/-0.13$  &$-0.06/+0.07$   & 0.08  \\ 
Ti I& $+0.12/-0.12$ &$+0.01/+0.01$  &$-0.06/+0.07$   & 0.08  \\
Ti II&$-0.04/+0.05$ &$+0.11/-0.14$  &$-0.14/+0.15$   & 0.11  \\
V I & $+0.14/-0.14$ &$+0.02/ 0.00$  &$-0.05/+0.07$   & 0.08  \\
Cr I& $+0.08/-0.09$ &$ 0.00/ 0.00$  &$-0.09/+0.08$   & 0.07  \\ 
Cr II&$-0.09/+0.09$ &$+0.10/-0.14$  &$-0.06/+0.06$   & 0.09  \\ 
Mn I& $-0.09/+0.09$ &$+0.01/-0.02$  &$-0.13/+0.13$   & 0.09  \\ 
Fe I& $+0.04/-0.04$ &$+0.02/-0.03$  &$-0.10/+0.10$   & 0.06  \\
Fe II&$-0.11/+0.12$ &$+0.13/-0.16$  &$-0.08/+0.10$   & 0.12  \\ 
Co I& $+0.05/-0.03$ &$+0.05/-0.04$  &$-0.05/+0.06$   & 0.05  \\ 
Ni I& $+0.02/-0.01$ &$+0.05/-0.06$  &$-0.08/+0.09$   & 0.06  \\ 
Cu I& $+0.01/-0.01$ &$+0.04/-0.05$  &$-0.06/+0.06$   & 0.04  \\
Zn I& $-0.07/+0.07$ &$+0.07/-0.10$  &$-0.15/+0.15$   & 0.11  \\
Y II& $-0.02/+0.01$ &$+0.11/-0.13$  &$-0.07/+0.08$   & 0.08  \\
Zr I& $+0.17/-0.19$ &$ 0.00/+0.01$  &$-0.02/+0.01$   & 0.10  \\
Ba II&$-0.01/+0.03$ &$+0.12/-0.14$  &$-0.21/+0.23$   & 0.15  \\
La II&$+0.01/ 0.00$ &$+0.11/-0.12$  &$-0.03/+0.05$   & 0.07  \\
Ce II&$+0.01/ 0.00$ &$+0.11/-0.12$  &$-0.02/+0.02$   & 0.07  \\
Nd II&$ 0.00/ 0.00$ &$+0.11/-0.12$  &$-0.03/+0.04$   & 0.07  \\
Sm II&$+0.01/ 0.00$ &$+0.11/-0.12$  &$-0.05/+0.07$   & 0.07  \\
Eu II&$-0.03/+0.02$ &$+0.10/-0.13$  &$-0.03/+0.02$   & 0.07  \\

\hline
 \end{tabular}
 \end{minipage}
\end{table} 

We evaluated the sensitivity of the derived abundances to the variations in adopted
atmospheric parameters by varying 
only one of the parameters by the amount corresponding to the typical error. 
The changes in abundances caused by varying atmospheric parameters by 100 K, 
0.25 cm s$^{-2}$ and 0.2 km s$^{-1}$ with respect to the chosen model atmosphere are
summarized in Table \ref{sensitivity}. 
We quadratically added the three contributors, by taking the square root of the sum
of the square of individual errors associated with uncertainties in temperature,
gravity and microturbulence, 
to obtain $\sigma_{2}$. The total error $\sigma_{tot}$ for each of the element is
the quadratic sum of $\sigma_{1}$ and $\sigma_{2}$. The error bars in the abundance
tables correspond to this total error.

\section{Results}

The four clusters support the
widely held impression that there is an abundance gradient such that the metallicity
[Fe/H] at the solar galactocentric distance decreases outwards (Magrini et al. 2009) at about -0.1 dex per
kpc: (R$_{gc}$, [Fe/H]) = (8.3, $-0.03$) for NGC 752, (9.3,$-0.07$) for NGC 2360,
(9.9, $-0.12$)  for NGC 1817, and (10.5,$-0.20$) for NGC 2506.

Results -- Tables 8--11 -- for the individual clusters are consistent 
with the assumption that stars within a
given cluster have the same composition.

If OCs are the principal supplier of field stars, there should be
a very close correspondence between the composition of stars in clusters and the
field. Such a correspondence represents a stiff challenge to the idea that
field stars have come from clusters because modern studies of field stars
show that there is no discernible `cosmic' dispersion in relative abundances --
[X/Fe] -- at a given [Fe/H] (Reddy et al. 2006). 
The four OCs are very likely representatives of the
Galactic thin disk but at their metallicity thin and thick disk stars very
likely have the same relative abundances. 

Several studies of thin disk dwarfs and giants have been reported
recently. For almost all elements over the [Fe/H] range sampled by these four OCs, the
field dwarfs and giants show a solar-like mix of elements, i.e., [X/Fe] $\simeq$ 0,
with very little star-to-star scatter at a given [Fe/H]. Sample papers echoing this
assertion
include Edvardsson et al. (1993), Bensby et al. (2005), Reddy et al. (2003, 2006),
Luck \& Heiter (2006) for dwarfs, and Mishenina et al. (2006), and Luck \& Heiter
(2007), and Takeda et al. (2008) for giants. These papers invoke, as we have done,
classical methods of abundance analysis involving
standard model atmospheres and LTE line formation.  

Methods of abundance analysis including choices of
gf-values, selection of model atmosphere grid and determination of solar reference
abundances differ among these
papers. Yet, the results suggest that differences of $\pm0.05$ and possibly
$\pm0.10$ dex
may arise among similar analyses by different authors of the same or similar stars.  
Such differences are attributable to measurement errors with the cosmic
dispersion masked by such errors.
One expects applications of the  classical method to
give slightly different results for dwarfs and giants for several reasons, e.g., the
effects of departures from LTE will be different for giants and dwarfs, and the
ability of standard atmospheres to represent true stellar atmospheres may differ for
dwarfs and giants. Thus, we restrict comparisons between our results and those
by similar methods for field giants i.e. systematic errors will be very similar
across this comparison.   

A useful comparison of abundances between our OCs and field giants may be
made using Luck \& Heiter's (2007) large sample of field giants analysed by methods
similar to ours, i.e., a differential analysis with respect to the Sun.
Using their Table 4, we calculated the mean abundances in field
giants across the [Fe/H] range of our clusters  (0.0 to $-0.2$) and those 
values are presented in column 6 of Table \ref{abundance}. Our 
cluster abundances in Table \ref{abundance} match the
abundances of the field giants to within 
about $\pm$0.15 dex, almost without exception.
The range of $\pm$0.15 dex assumes measurement uncertainties of
about 0.1 dex in both studies. (Luck \& Heiter did not include Zr and Sm in their
collection of elements.).      
Luck \& Heiter's results for field giants are generally confirmed by
Mishenina et al.'s (2006)
and Takeda et al.'s (2008) for other large samples of field giants. One may
note that Takeda et al.'s Mn, Ce, and Nd abundances  (relative to Fe) agree well with
ours but their analysis while it gives ionization equilibrium for Fe does not
do so for Sc, Ti, V, and Cr.

\begin{table*}
 \begin{minipage}{120mm}
 \caption{Elemental abundances for the four clusters in this study and the thin disk mean values from
          Luck \& Heiter (2007) in the metallicity range of our clusters (i.e. 0.0 to $-$0.2).}
 \label{abundance}
 \begin{tabular}{lllllc}
  \hline
\multicolumn{1}{l}{Species}&  
\multicolumn{1}{l}{NGC 752}&  
\multicolumn{1}{l}{NGC 1817}&  
\multicolumn{1}{l}{NGC 2360}&  
\multicolumn{1}{l}{NGC 2506} &
\multicolumn{1}{c}{Thin Disk} \\
\hline

$[$Na I/Fe$]$  &$+0.12\pm0.03$ &$+0.16\pm0.02$ &$+0.20\pm0.03$  &$+0.21\pm0.03$ &$0.10\pm0.06$ \\
$[$Mg I/Fe$]$  &$-0.01\pm0.03$ &$+0.08\pm0.03$ &$+0.07\pm0.03$  &$+0.05\pm0.04$ &$0.08\pm0.10$ \\
$[$Al I/Fe$]$  &$+0.15\pm0.02$ &$+0.11\pm0.02$ &$+0.09\pm0.02$  &$+0.17\pm0.01$ &$0.09\pm0.05$ \\
$[$Si I/Fe$]$  &$+0.11\pm0.02$ &$+0.10\pm0.02$ &$+0.15\pm0.02$  &$+0.04\pm0.02$ &$0.12\pm0.04$ \\
$[$Ca I/Fe$]$  &$+0.03\pm0.05$ &$+0.14\pm0.04$ &$+0.11\pm0.04$  &$+0.10\pm0.05$ &$-0.04\pm0.05$ \\
$[$Sc I/Fe$]$  &$+0.07\pm0.04$ &               &$+0.06\pm0.04$  &$ 0.00\pm0.02$ &$-0.08\pm0.06$ \\
$[$Sc II/Fe$]$ &$+0.04\pm0.04$ &$ 0.00\pm0.04$ &$+0.03\pm0.04$  &$+0.05\pm0.05$ & \\ 
$[$Ti I/Fe$]$  &$-0.07\pm0.05$ &$ 0.00\pm0.04$ &$-0.03\pm0.05$  &$+0.04\pm0.05$ &$0.00\pm0.03$ \\
$[$Ti II/Fe$]$ &$-0.04\pm0.05$ &$+0.02\pm0.05$ &$-0.05\pm0.05$  &$+0.03\pm0.05$ & \\
$[$V I/Fe$]$   &$+0.03\pm0.05$ &$+0.01\pm0.04$ &$+0.08\pm0.04$  &$+0.01\pm0.05$ &$-0.09\pm0.07$\\
$[$V II/Fe$]$  &$+0.04\pm0.04$ &               &$-0.05\pm0.03$  &               & \\
$[$Cr I/Fe$]$  &$-0.03\pm0.04$ &$ 0.00\pm0.03$ &$+0.01\pm0.04$  &$-0.01\pm0.04$ &$+0.01\pm0.05$ \\ 
$[$Cr II/Fe$]$ &$+0.02\pm0.05$ &$+0.03\pm0.04$ &$ 0.00\pm0.05$  &$-0.09\pm0.04$ & \\ 
$[$Mn I/Fe$]$  &  $\bf-0.13 $  &   $\bf-0.18$  &   $\bf-0.21$   &$\bf-0.18$     &$+0.06\pm0.07$ \\
$[$Fe I/H$]$   &$-0.04\pm0.03$ &$-0.13\pm0.04$ &$-0.08\pm0.03$  &$-0.22\pm0.04$ & \\
$[$Fe II/H$]$  &$-0.02\pm0.05$ &$-0.11\pm0.05$ &$-0.07\pm0.06$  &$-0.19\pm0.06$ & \\ 
$[$Co I/Fe$]$  &$-0.02\pm0.02$ &$+0.03\pm0.03$ &$+0.06\pm0.03$  &$-0.02\pm0.03$ &$+0.06\pm0.08$ \\ 
$[$Ni I/Fe$]$  &$-0.01\pm0.03$ &$-0.02\pm0.03$ &$+0.01\pm0.03$  &$-0.08\pm0.03$ &$0.00\pm0.03$ \\ 
$[$Cu I/Fe$]$  & $\bf-0.11$    &   $\bf-0.23$  &   $\bf-0.18$   &$\bf-0.12$     &$+0.01\pm0.13$ \\
$[$Zn I/Fe$]$  &$-0.10\pm0.04$ & $0.00\pm0.05$ &$+0.04\pm0.08$  &$+0.01\pm0.05$ &$ $ \\  
$[$Y II/Fe$]$  &$+0.03\pm0.03$ &$+0.07\pm0.05$ &$+0.06\pm0.04$  &$+0.04\pm0.07$ &$+0.07\pm0.15$ \\
$[$Zr I/Fe$]$  &$+0.06\pm0.05$ &$+0.08\pm0.05$ &$+0.08\pm0.05$  &$ $            & \\
$[$Ba II/Fe$]$ & $\bf+0.13$    &   $\bf+0.13$  &   $\bf+0.10$   &$\bf+0.31$     &$+0.04\pm0.16$ \\
$[$La II/Fe$]$ &$+0.13\pm0.03$ &$+0.12\pm0.03$ &$+0.14\pm0.05$  &$+0.28\pm0.04$ &$+0.05\pm0.09$ \\
$[$Ce II/Fe$]$ &$\bf+0.13$     &$\bf+0.20$     &$\bf+0.18 $     &$\bf+0.18$     &$+0.05\pm0.09$ \\  
$[$Nd II/Fe$]$ &$+0.06\pm0.04$ &$+0.14\pm0.04$ &$+0.06\pm0.04$  &$+0.16\pm0.06$ &$-0.01\pm0.07$ \\
$[$Sm II/Fe$]$ &$\bf+0.08$     &$\bf+0.21 $    &$\bf+0.13 $     &$\bf+0.22$     & \\
$[$Eu~II/Fe$]$ & $\bf+0.07$    &  $\bf+0.13$   &$\bf+0.04$      &$\bf+0.22$     &$0.08\pm0.06$ \\

\hline
\end{tabular}
 \flushbottom{{\bf Note}: Abundances calculated by synthesis are presented in bold numbers.}
\end{minipage}
\end{table*}

Close scrutiny of our and Luck \& Heiter's 
abundances suggest two possible 
differences: (i) the OCs appear to have a low [Mn/Fe] than local field giants; 
(ii) the OCs relative to the field giants may be enriched in Ba and heavier elements.

The [Mn/Fe] ratio decreases with [Fe/H], as shown by Luck \& Heiter, and others. If one 
takes into account the decrease found for field giants, the [Mn/Fe] for the OCs is on 
average 0.12 dex lower than for the field giants. We suppose that this offset is not 
implausibly considered to be a systematic error arising from two similar but identical 
analyses\footnote{McWilliam et al. (2003) reported low [Mn/Fe] at [Fe/H] $\simeq$0 
for giants in the Galactic bulge but such stars are also enriched in the $\alpha$- elements
and Eu, characteristics not carried by the giants in our OCs.}.

Abundances for the heaviest elements are based on either strong lines (e.g., Ba) or on 
just one to three lines. Thus, the differences between OC and field giants may
be in part due to systematic errors. However, D'Orazi et al. (2009) and Maiorca 
et al. (2011) concluded that heavy element abundances increase from old to young clusters.

\section{Conclusions} 
This study is a part of a project on the determination of chemical abundances of
OCs through high resolution spectroscopy, whose final goal is to chemically tag the
disk field stars to re-construct the dispersed stellar aggregates and to derive the
abundance gradient in the Galactic disk.        
In this paper, we have presented an analysis 
of giant stars in four OCs located between Rgc $\sim$ 8.3 to 10.5 kpc. 

The main results of our study are: 

\begin{enumerate}
 \item Membership of the giants used for abundance analysis in each of the
clusters has been confirmed through their radial velocities; 
 \item Based on high-resolution spectra and standard LTE analysis, we have derived
stellar parameters and abundance ratios of the light elements (Na, Al),
$\alpha$-elements (Mg, Si, Ca, Ti), iron-peak elements (Sc, V, Cr, Mn, Fe, Co, Ni),
light s-process elements (Y, Zr), heavy s-process elements (Ba, La, Ce, Nd), and of
the r-process elements (Sm, Eu).
\item We have derived average [Fe/H] values of $-$0.02$\pm$0.05 for
NGC 752, $-$0.07$\pm$0.06 for NGC 2360, $-$0.11$\pm$0.05 for NGC 1817 and
$-$0.19$\pm$0.06 NGC 2506. 
\item Comparison of our results with published abundances for thin disk field
 giants show very similar chemical compositions. Field and OC giants with 
 [Fe/H] $\sim$ 0 have identical compositions to within the errors of measurements. 
\item Two hints that the OC and field giant compositions are not precisely the 
 same at the same metallicity (i.e., [Fe/H] = 0) will be examined in future papers.
 These hints are a Mn deficiencies and a heavy element overabundance in the OCs.
\end{enumerate}

\vskip3ex \noindent
{\bf Acknowledgments:}
DLL wishes to thank the Robert A. Welch Foundation of Houston, Texas for support
throug grant F-634.

\begin{table*}
\centering
 \begin{minipage}{145mm}
 \caption{Elemental abundances for stars in the OC NGC 752.}
 \label{NGC752}
 \begin{tabular}{llllll}
  \hline
\multicolumn{1}{c}{Species}&  
\multicolumn{1}{c}{star no. 77}&  
\multicolumn{1}{c}{star no. 137}&  
\multicolumn{1}{c}{star no. 295}&  
\multicolumn{1}{c}{star no. 311}&
\multicolumn{1}{c}{Average} \\
\hline

$[$Na I/Fe$]$ &$+0.13\pm0.05$(3) &$+0.10\pm0.03$(3) &$+0.14\pm0.02$(3)
&$+0.13\pm0.07$(3)&$+0.12\pm0.02$ \\
$[$Mg I/Fe$]$ &$-0.01\pm0.07$(3) &$+0.02\pm0.05$(2) &$-0.06\pm0.05$(3)
&$-0.01\pm0.04$(3)&$-0.01\pm0.03$ \\
$[$Al I/Fe$]$ &$+0.19\pm0.04$(3) &$+0.11\pm0.03$(4) &$+0.14\pm0.04$(3)
&$+0.17\pm0.02$(3)&$+0.15\pm0.02$ \\
$[$Si I/Fe$]$
&$+0.12\pm0.05$(13)&$+0.03\pm0.05$(10)&$+0.11\pm0.07$(12)&$+0.18\pm0.06$(14)&$+0.11\pm0.03$
\\
$[$Ca I/Fe$]$ &$+0.01\pm0.07$(8) &$+0.03\pm0.05$(7) &$+0.06\pm0.05$(7)
&$+0.01\pm0.06$(8)&$+0.03\pm0.03$ \\
$[$Sc I/Fe$]$ &$+0.07\pm0.04$(6) &$+0.06\pm0.10$(3) &$+0.12\pm0.03$(5)
&$+0.03\pm0.05$(4)&$+0.07\pm0.03$ \\
$[$Sc II/Fe$]$&$-0.03\pm0.04$(6) &$+0.19\pm0.03$(5) &$+0.01\pm0.05$(5)
&$+0.01\pm0.03$(6)&$+0.04\pm0.02$ \\ 
$[$Ti I/Fe$]$
&$-0.03\pm0.06$(13)&$-0.14\pm0.04$(15)&$-0.02\pm0.06$(13)&$-0.08\pm0.06$(13)&$-0.07\pm0.03$
\\
$[$Ti II/Fe$]$&$-0.06\pm0.08$(6) &$-0.06\pm0.09$(5) &$ 0.00\pm0.09$(7)
&$-0.03\pm0.06$(7)&$-0.04\pm0.04$ \\
$[$V I/Fe$]$ 
&$+0.11\pm0.06$(10)&$-0.05\pm0.07$(6)&$+0.05\pm0.04$(12)&$+0.03\pm0.05$(10)&$+0.03\pm0.03$
\\
$[$V II/Fe$]$ &$ 0.00\pm0.07$(3) &$               $ &$+0.04\pm0.08$(3)
&$+0.07\pm0.08$(3)&$+0.04\pm0.04$ \\
$[$Cr I/Fe$]$ &$-0.02\pm0.05$(12)&$-0.07\pm0.04$(11)&$-0.04\pm0.07$(11)&$
0.00\pm0.06$(11)&$-0.03\pm0.03$ \\ 
$[$Cr II/Fe$]$ &$+0.06\pm0.05$(4) &$-0.03\pm0.06$(6) &$+0.01\pm0.02$(4)
&$+0.04\pm0.02$(4)&$+0.02\pm0.02$ \\ 
$[$Mn I/Fe$]$ &$\bf-0.11$        &$\bf-0.18$        &$-0.12$           &$\bf-0.10$  
    &$\bf-0.13 $      \\
$[$Fe I/H $]$ &$-0.04\pm0.05$(48)&$-0.02\pm0.06$(46)&$-0.05\pm0.05$(43)
&$-0.04\pm0.05$(43)&$-0.04\pm0.03$ \\
$[$Fe II/H$]$ &$-0.02\pm0.05$(13)&$ 0.00\pm0.05$(12) &$-0.04\pm0.06$(13)
&$-0.04\pm0.04$(13)&$-0.02\pm0.02$ \\ 
$[$Co I/Fe$]$ &$-0.03\pm0.05$(5) &$-0.08\pm0.05$(7) &$ 0.00\pm0.08$(6)
&$+0.04\pm0.06$(5)&$-0.02\pm0.03$   \\ 
$[$Ni I/Fe$]$ &$+0.01\pm0.03$(12) &$-0.04\pm0.06$(11)&$-0.01\pm0.05$(14) &$
0.00\pm0.05$(13)&$-0.01\pm0.02$ \\ 
$[$Cu I/Fe$]$ &   $\bf -0.03 $   &  $\bf-0.12$      &$\bf-0.15$   &$\bf-0.13$  
    &$\bf-0.11$    \\
$[$Zn I/Fe$]$ & $-0.09\pm0.00$  &               &$-0.07\pm0.00$    &$-0.14\pm0.00$ 
&$-0.10\pm0.00$     \\
$[$Y II/Fe$]$ &$+0.03\pm0.01$(3) &$-0.01\pm0.04$(6) &$+0.06\pm0.02$(4)
&$+0.03\pm0.03$(4)&$+0.03\pm0.01$ \\
$[$Zr I/Fe$]$  &$+0.06\pm0.01$(3) &$-0.05\pm0.01$(3) &$+0.16\pm0.02$(3)
&$+0.06\pm0.01$(4)&$+0.06\pm0.01$ \\
$[$Ba II/Fe$]$ &  $\bf +0.14 $    &    $\bf+0.12$    &$\bf+0.16$        &$\bf+0.11$ 
     &$\bf+0.13$       \\ 
$[$La II/Fe$]$ &$+0.15\pm0.06$(2) &$+0.07\pm0.08$(2) &$+0.14\pm0.06$(2) &$+0.16\pm0.06$(2)
 &$+0.13\pm0.03$ \\
$[$Ce II/Fe$]$ &$\bf+0.13 $  &$\bf+0.08$  &$\bf+0.14$ &$\bf+0.16$ &$\bf+0.13$ \\
$[$Nd II/Fe$]$ &$+0.07\pm0.09$(3) &$+0.02\pm0.04$(4) &$+0.06\pm0.03$(3)
&$+0.10\pm0.04$(4)&$+0.06\pm0.03$ \\
$[$Sm II/Fe$]$ &$\bf+0.08$ &                  &$\bf+0.07$ &$\bf+0.11$ &$\bf+0.08$ \\
$[$Eu II/Fe$]$ &  $\bf +0.10 $    & $\bf+0.00$       &$\bf+0.09$        &$\bf+0.08$ 
     &$\bf+0.07$    \\

\hline
 \end{tabular}
\vskip1ex
\flushbottom{{\bf Note}: The abundances calculated by synthesis are presented in
bold numbers.
The remaining elemental abundances were calculated using line equivalent widths.
Numbers in the parentheses indicate the number of lines used in calculating 
the abundance of that element.
In this analysis we have adopted the hfs data of Prochaska \& McWilliam (2000) for
Mn I, Mucciarelli et al. (2008) for Eu II line, McWilliam (1998) 
for Ba II line, and Allen et al. (2011) for Cu I lines.
}
\end{minipage}
\end{table*}

\begin{table*}
\centering
 \begin{minipage}{145mm}
 \caption{Elemental abundances for stars in the OC NGC 1817.}
 \label{NGC1817}
 \begin{tabular}{lllll}
  \hline
\multicolumn{1}{c}{Species}&  
\multicolumn{1}{c}{star no. 1027}&  
\multicolumn{1}{c}{star no. 2038}&  
\multicolumn{1}{c}{star no. 2059}&  
\multicolumn{1}{c}{Average} \\
\hline

$[$Na I/Fe$]$ &$+0.15\pm0.04$(3) &$+0.23\pm0.03$(4) &$+0.09\pm0.04$(4)
&$+0.16\pm0.02$    \\
$[$Mg I/Fe$]$ &$+0.03\pm0.03$(3) &$+0.12\pm0.06$(4) &$+0.08\pm0.01$(2)
&$+0.08\pm0.02$    \\
$[$Al /Fe$]$I &$+0.10\pm0.03$(3) &$+0.12\pm0.03$(2) &$+0.12\pm0.06$(2)
&$+0.11\pm0.02$    \\
$[$Si I/Fe$]$
&$+0.05\pm0.06$(12)&$+0.12\pm0.06$(10)&$+0.14\pm0.06$(10)&$+0.10\pm0.03$    \\
$[$Ca I/Fe$]$ &$+0.19\pm0.04$(9) &$+0.20\pm0.05$(9) &$+0.04\pm0.04$(9)
&$+0.14\pm0.02$    \\
$[$Sc II/Fe$]$&$ 0.00\pm0.05$(5) &$-0.04\pm0.08$(5) &$+0.04\pm0.06$(5) &$
0.00\pm0.04$    \\ 
$[$Ti I/Fe$]$ &$+0.07\pm0.05$(8)&$+0.04\pm0.06$(7)&$-0.11\pm0.03$(8)&$ 0.00\pm0.03$
     \\
$[$Ti II/Fe$]$&$+0.09\pm0.04$(9) &$-0.06\pm0.06$(9) &$+0.03\pm0.05$(9)
&$+0.02\pm0.03$    \\
$[$V I/Fe$]$ 
&$+0.06\pm0.04$(8)&$-0.01\pm0.05$(8)&$-0.01\pm0.07$(10)&$+0.01\pm0.03$    \\
$[$Cr I/Fe$]$ &$-0.02\pm0.05$(11)&$-0.02\pm0.05$(11)&$+0.03\pm0.06$(13)&$
0.00\pm0.03$    \\ 
$[$Cr II/Fe$]$&$-0.01\pm0.05$(6) &$+0.01\pm0.05$(4) &$+0.08\pm0.06$(5)
&$+0.03\pm0.03$    \\ 
$[$Mn I/Fe$]$ &$\bf-0.17 $       &$\bf-0.14 $       &$\bf-0.24 $      &$\bf-0.18$  
    \\
$[$Fe I/H $]$
&$-0.15\pm0.05$(30)&$-0.14\pm0.05$(43)&$-0.11\pm0.05$(45)&$-0.13\pm0.03$    \\
$[$Fe II/H$]$ &$-0.11\pm0.06$(11) &$-0.14\pm0.06$(11) &$-0.09\pm0.06$(11)
&$-0.11\pm0.03$    \\ 
$[$Co I/Fe$]$ &$+0.03\pm0.08$(5) &$+0.10\pm0.05$(4) &$-0.03\pm0.07$(5)
&$+0.03\pm0.04$    \\ 
$[$Ni I/Fe$]$ &$
0.00\pm0.03$(11)&$-0.03\pm0.05$(13)&$-0.02\pm0.05$(12)&$-0.02\pm0.02$    \\ 
$[$Cu I/Fe$]$ &   $\bf -0.22 $    &  $\bf-0.23$     &$\bf-0.24$       &$\bf-0.23$  
    \\
$[$Zn I/Fe$]$ &$-0.06\pm0.00$(1) &$+0.11\pm0.00$(1) &$-0.07\pm0.00$(1) &$
0.00\pm0.00 $   \\ 
$[$Y II/Fe$]$ &$+0.10\pm0.08$(5) &$+0.03\pm0.07$(5) &$+0.07\pm0.03$(4)
&$+0.07\pm0.04$    \\
$[$Zr I/Fe$]$ &$+0.14\pm0.00$(1) &$+0.11\pm0.08$(2) &$-0.01\pm0.06$(2)
&$+0.08\pm0.03$    \\
$[$Ba II/Fe$]$ &  $ \bf+0.16 $    &  $\bf+0.11$     &$\bf+0.12$        &$\bf+0.13$   \\
$[$La II/Fe$]$ &$+0.16\pm0.06$(1) &$+0.09\pm0.06$(1) &$+0.12\pm0.06$(2) &$+0.12\pm0.03$ \\
$[$Ce II/Fe$]$ &$\bf+0.22$  &   $\bf+0.19 $       &$\bf+0.18$ &$\bf+0.20$    \\
$[$Nd II/Fe$]$ &$+0.14\pm0.07$(4) &$+0.17\pm0.03$(3) &$+0.12\pm0.05$(3)
&$+0.14\pm0.03$    \\
$[$Sm II/Fe$]$ &$\bf+0.22$     &$  $  &$\bf+0.21$ &$\bf+0.21$    \\
$[$Eu II/Fe$]$ &  $\bf +0.16 $    & $ $       &$\bf+0.11$         &$\bf+0.13$       
    \\

\hline
 \end{tabular}
\vskip1ex
\flushbottom{{\bf Note}: Same as in table 8. }
\end{minipage}
\end{table*}

\begin{table*}
\centering  
 \begin{minipage}{145mm}
 \caption{Elemental abundances for stars in the OC NGC 2506.}
 \label{NGC2506}
 \begin{tabular}{lllll}
  \hline
\multicolumn{1}{c}{Species}&  
\multicolumn{1}{c}{star no. 2212}&  
\multicolumn{1}{c}{star no. 3231}&  
\multicolumn{1}{c}{star no. 4138}&  
\multicolumn{1}{c}{Average} \\
\hline

$[$Na I/Fe$]$ &$+0.18\pm0.09$(3) &$+0.14\pm0.08$(5) &$+0.32\pm0.03$(3)
&$+0.21\pm0.04$   \\
$[$Mg I/Fe$]$ &$+0.19\pm0.03$(3) &$+0.03\pm0.07$(3) &$-0.08\pm0.08$(2)
&$+0.05\pm0.04$   \\
$[$Al I/Fe$]$ &$+0.11\pm0.01$(2) &$+0.19\pm0.03$(2) &$+0.21\pm0.03$(2)
&$+0.17\pm0.01$   \\
$[$Si I/Fe$]$
&$+0.03\pm0.09$(12)&$+0.03\pm0.08$(7)&$+0.06\pm0.06$(10)&$+0.04\pm0.04$   \\
$[$Ca I/Fe$]$ &$+0.09\pm0.09$(9) &$+0.09\pm0.09$(9) &$+0.11\pm0.06$(8)
&$+0.10\pm0.05$  \\
$[$Sc I/Fe$]$ &$+0.07\pm0.00$(1) &$-0.01\pm0.00$(1) &$-0.06\pm0.00$(1) &$
0.00\pm0.00$   \\
$[$Sc II/Fe$]$&$-0.01\pm0.08$(5) &$+0.09\pm0.09$(5) &$+0.08\pm0.07$(3)
&$+0.05\pm0.05$   \\ 
$[$Ti I /Fe$]$&$-0.08\pm0.07$(9) &$-0.01\pm0.09$(9) &$+0.17\pm0.07$(6)
&$+0.04\pm0.04$   \\
$[$Ti II/Fe$]$&$-0.07\pm0.07$(9) &$+0.13\pm0.06$(6) &$+0.04\pm0.08$(8)
&$+0.03\pm0.04$   \\
$[$V I /Fe$]$ &$-0.14\pm0.08$(9) &$+0.05\pm0.07$(8) &$+0.12\pm0.08$(8)
&$+0.01\pm0.04$   \\
$[$Cr I/Fe$]$ &$-0.02\pm0.07$(6) &$-0.02\pm0.08$(11)&$ 0.00\pm0.04$(8)
&$-0.01\pm0.04$   \\ 
$[$Cr II/Fe$]$&$-0.14\pm0.07$(6) &$-0.04\pm0.06$(6) &$-0.08\pm0.09$(4)
&$-0.09\pm0.04$   \\ 
$[$Mn I/Fe$]$ &$\bf-0.21$        &$\bf-0.16 $       &$\bf-0.16 $        &$\bf-0.18$   \\
$[$Fe I/H $]$
&$-0.19\pm0.06$(33)&$-0.25\pm0.06$(38)&$-0.21\pm0.05$(31)&$-0.22\pm0.03$   \\
$[$Fe II/H$]$ &$-0.17\pm0.05$(8) &$-0.22\pm0.06$(8) &$-0.19\pm0.07$(9)
&$-0.19\pm0.03$   \\ 
$[$Co I/Fe$]$ &$-0.02\pm0.07$(5) &$-0.05\pm0.09$(4) &$+0.02\pm0.05$(3)
&$-0.02\pm0.04$   \\ 
$[$Ni I/Fe$]$
&$-0.08\pm0.06$(12)&$-0.09\pm0.08$(10)&$-0.07\pm0.05$(11)&$-0.08\pm0.04$   \\ 
$[$Cu I/Fe$]$ &   $\bf -0.10 $   &  $\bf-0.10$      &$\bf-0.15$         &$\bf-0.12$
   \\
$[$Zn I/Fe$]$ &$-0.01\pm0.00$(1) &$+0.05\pm0.00$(1) &$ 0.00\pm0.00$(1)
&$+0.01\pm0.00$   \\ 
$[$Y II/Fe$]$ &$+0.07\pm0.12$(1)&$+0.03\pm0.12$(1) &$+0.01\pm0.12$(1) &$+0.04\pm0.07$   \\
$[$Ba II/Fe$]$ &  $\bf +0.29$    &    $\bf+0.31$     & $\bf+0.31$        &$\bf+0.31$
   \\
$[$La II/Fe$]$ &$+0.21\pm0.09$(2) &$+0.31\pm0.07$(1) &$+0.31\pm0.07$(1)  &$+0.28\pm0.04$ \\
$[$Ce II/Fe$]$ &$\bf+0.14$    &$ $  &$\bf+0.23$  &$\bf+0.18$   \\
$[$Nd II/Fe$]$ &$ $  &$ $  &$+0.16\pm0.09$(3) &$+0.16\pm0.09$   \\
$[$Sm II/Fe$]$ &$\bf+0.24$  &          &$\bf+0.21$   &$\bf+0.22$   \\
$[$Eu II/Fe$]$ &  $\bf+0.17 $    & $\bf+0.26$       &$\bf+0.23$         &$\bf+0.22$ 
  \\

\hline
 \end{tabular}
 \vskip1ex
\flushbottom{{\bf Note}: Same as in table 8. }
\end{minipage}
\end{table*}

\begin{table*}
\centering
 \begin{minipage}{145mm}
 \caption{Elemental abundances for stars in the OC NGC 2360.}
 \label{NGC2360}
 \begin{tabular}{llllll}
  \hline
\multicolumn{1}{c}{Species}&  
\multicolumn{1}{c}{star no. 5}&  
\multicolumn{1}{c}{star no. 6}&  
\multicolumn{1}{c}{star no. 8}&  
\multicolumn{1}{c}{star no. 12}&
\multicolumn{1}{c}{Average} \\
\hline

$[$Na I/Fe$]$  &$+0.14\pm0.07$(3) &$+0.23\pm0.06$(4) &$+0.21\pm0.05$(3)
&$+0.24\pm0.04$(4)&$+0.20\pm0.03$  \\
$[$Mg I/Fe$]$  &$-0.06\pm0.01$(3) &$+0.13\pm0.07$(4) &$+0.13\pm0.08$(4)
&$+0.10\pm0.04$(2)&$+0.07\pm0.03$  \\
$[$Al I/Fe$]$  &$+0.18\pm0.06$(2) &$ 0.00\pm0.01$(3) &$+0.19\pm0.05$(3)
&$-0.01\pm0.05$(3)&$+0.09\pm0.02$  \\
$[$Si I/Fe$]$ 
&$+0.14\pm0.06$(15)&$+0.18\pm0.06$(14)&$+0.14\pm0.05$(11)&$+0.16\pm0.06$(5)&$+0.15\pm0.03$
 \\
$[$Ca I/Fe$]$  &$+0.03\pm0.05$(8) &$+0.14\pm0.06$(9)&$+0.16\pm0.04$(8)
&$+0.13\pm0.06$(11)&$+0.11\pm0.03$ \\
$[$Sc I/Fe$]$  &$+0.19\pm0.01$(2) &$+0.14\pm0.06$(4) &$             $  
&$-0.10\pm0.06$(3) &$+0.06\pm0.03$  \\
$[$Sc II/Fe$]$ &$ 0.00\pm0.03$(6) &$+0.15\pm0.07$(5) &$+0.05\pm0.06$(5)
&$-0.09\pm0.03$(5) &$+0.03\pm0.02$  \\ 
$[$Ti I/Fe$]$  &$+0.04\pm0.04$(13)
&$+0.02\pm0.07$(12)&$+0.01\pm0.05$(11)&$-0.18\pm0.07$(13)&$-0.03\pm0.03$  \\
$[$Ti II/Fe$]$ &$+0.07\pm0.02$(2) &$ 0.00\pm0.07$(8) &$ 0.00\pm0.05$(7)
&$-0.10\pm0.02$(7) &$ 0.00\pm0.02$  \\
$[$V I /Fe$]$ 
&$+0.23\pm0.05$(12)&$+0.04\pm0.06$(9)&$+0.06\pm0.05 $ &$-0.01\pm0.07$(5)
&$+0.08\pm0.03$  \\
$[$V II/Fe$]$  &               &$-0.05\pm0.09$(3) &$-0.06\pm0.09$(2)  &            
  &$-0.05\pm0.02$  \\
$[$Cr I/Fe$]$  &$
0.00\pm0.05$(11)&$+0.02\pm0.06$(14)&$+0.02\pm0.03$(10)&$+0.02\pm0.05$(13)&$+0.01\pm0.02$
 \\ 
$[$Cr II/Fe$]$ &$+0.02\pm0.08$(4) &$+0.01\pm0.06$(6) &$-0.04\pm0.03$(6)
&$+0.02\pm0.06$(7) &$ 0.00\pm0.03$  \\ 
$[$Mn I/Fe$]$  &$\bf-0.14$        &$\bf-0.18 $       &$\bf-0.20 $        &$ $
     &$\bf-0.21$    \\
$[$Fe I/H $]$ 
&$-0.07\pm0.07$(54)&$-0.13\pm0.05$(32)&$-0.05\pm0.06$(43)&$-0.06\pm0.07$(60)&$-0.08\pm0.03$
 \\
$[$Fe II/H$]$
&$-0.07\pm0.04$(11)&$-0.10\pm0.04$(10)&$-0.06\pm0.06$(10)&$-0.04\pm0.05$(9)&$-0.07\pm0.02$
 \\ 
$[$Co I/Fe$]$  &$+0.17\pm0.07$(7) &$ 0.00\pm0.07$(6) &$+0.05\pm0.06$(6)
&$+0.03\pm0.09$(9) &$+0.06\pm0.04$  \\ 
$[$Ni I/Fe$]$ 
&$-0.02\pm0.06$(15)&$+0.01\pm0.04$(13)&$+0.02\pm0.06$(14)&$+0.04\pm0.08$(16)&$+0.01\pm0.03$
 \\ 
$[$Cu I/Fe$]$  &   $\bf -0.20 $    &  $\bf-0.16$       &$\bf-0.23$        
&$\bf-0.13$    &$\bf-0.18$   \\
$[$Zn I/Fe$]$  &$ $  &  $+0.05\pm0.00 $(1)&$ $ 
&$+0.03\pm0.00$(1)&$+0.04\pm0.08$    \\ 
$[$Y II/Fe$]$  &$+0.14\pm0.03$(3) &$+0.08\pm0.05$(4) &$-0.04\pm0.06$(4)
&$+0.07\pm0.06$(5) &$+0.06\pm0.02$  \\
$[$Zr I/Fe$]$  &$+0.13\pm0.02$(3) &$+0.16\pm0.01$(2) &$+0.10\pm0.06$(4)
&$-0.06\pm0.06$(3) &$+0.08\pm0.02$  \\
$[$Ba II/Fe$]$ &  $\bf +0.06 $    &    $\bf+0.13$    &$\bf+0.11$         &$  $      
 &$\bf+0.10$     \\
$[$La II/Fe$]$ &$ $  &$+0.17\pm0.10$(2) &$+0.17\pm0.09$(2) &$+0.08\pm0.09$(2) &$+0.14\pm0.05$ \\
$[$Ce II/Fe$]$ &$+0.18$  &$ +0.19   $  &$ $ &$  $ &$+0.18 $  \\
$[$Nd II/Fe$]$ &$ $  &$+0.07\pm0.05$(4) &$+0.07\pm0.06$(7)
&$+0.07\pm0.05$(3) &$+0.06\pm0.03$  \\
$[$Sm II/Fe$]$ &$\bf+0.13$ &         &$\bf+0.12$ & $\bf+0.13$  &$+0.13$  \\
$[$Eu II/Fe$]$ &  $     $    & $\bf+0.05$       &$\bf+0.04$         &$\bf+0.02$
      &$\bf+0.04$      \\

\hline
 \end{tabular}
 \vskip1ex
\flushbottom{{\bf Note}: Same as in table 8. }
\end{minipage}
\end{table*}

\begin{table*}
 \caption{The linelist for all program stars from each of the OCs presented in this paper.}
 \label{EWmeasurement}
 \begin{tabular}{lllllllllllccccccc}
 \hline

& &    &    & {EW(m\AA) for NGC 752} & {EW(m\AA) for NGC 1817} &
    {EW(m\AA) for NGC 2360} & {EW(m\AA) for NGC 2506} \\  
$\lambda$(\AA) & Species$^{a}$ & LEP$^{b}$ & $\log~{gf}$ & {$\#$77\hspace{0.1cm} $\#$137\hspace{0.1cm} $\#$295\hspace{0.1cm} $\#$311} & {$\#$1027\hspace{0.05cm} $\#$2038\hspace{0.05cm} $\#$2059}
& {$\#$5\hspace{0.25cm} $\#$6\hspace{0.25cm} $\#$8\hspace{0.25cm} $\#$12} & $\#$2212 $\#$3231 $\#$4138  \\
\hline

4668.53 & 11.0 & 2.10 & -1.25& - - -\hspace{0.3cm} - - -\hspace{0.3cm} - - -\hspace{0.3cm}- - -  & - - -\hspace{0.45cm} 79.6\hspace{0.35cm}- - -  & - - - \hspace{0.05cm} 84.6\hspace{0.1cm} 97.0\hspace{0.2cm}- - - & 85.3\hspace{0.32cm}  75.2\hspace{0.35cm}  85.2 \\
4982.79 & 11.0 & 2.10 & -0.91& - - - \hspace{0.15cm} 111.6\hspace{0.2cm} - - -\hspace{0.2cm} - - -   & 99.4\hspace{0.45cm}104.2\hspace{0.25cm}107.2& - - -\hspace{0.3cm}- - -\hspace{0.15cm} - - -\hspace{0.10cm} 121.3 &- - -\hspace{0.35cm} 96.3\hspace{0.2cm} 102.8  \\
5688.21 & 11.0 & 2.10 & -0.45&\hspace{-0.2cm}145.9 \hspace{0.22cm}140.7\hspace{0.2cm}151.1\hspace{0.2cm} - - - & - - -\hspace{0.5cm}- - -\hspace{0.5cm}- - - & \hspace{-0.2cm}141.2   143.4   140.7\hspace{0.08cm}  155.1& - - -\hspace{0.5cm}- - -\hspace{0.5cm}- - - \\
6154.23 & 11.0 & 2.10 & -1.55& 74.5 \hspace{0.2cm} 75.4\hspace{0.2cm} 69.9\hspace{0.2cm} 75.5& 59.5\hspace{0.45cm} 64.6\hspace{0.35cm}70.8&\hspace{-0.2cm} 67.6\hspace{0.2cm} 70.1\hspace{0.2cm}68.9\hspace{0.2cm} 91.3 &  70.2 \hspace{0.2cm}  50.3\hspace{0.4cm} - - -   \\
6160.76 & 11.0 & 2.10 & -1.25& 96.1 \hspace{0.2cm} 98.5\hspace{0.2cm} 89.5\hspace{0.2cm} 96.3& 83.5\hspace{0.45cm} 85.4\hspace{0.35cm}93.0&\hspace{-0.15cm} 94.6\hspace{0.15cm} 91.7\hspace{0.2cm}91.4 \hspace{0.2cm}- - - &  100.0 \hspace{0.1cm}  81.7\hspace{0.4cm}- - - \\

\hline
 \end{tabular}
 \vskip1ex
\flushleft{{\bf Notes}:   \\
 $^{a}$ The integer part of the 'Species' indicates the atomic number, and the decimal 
  component indicates the ionization state \\ \hspace{0.2cm} (0 = neutral, 1 = singly ionized). \\ \vskip1ex
 $^{b}$ All the lines are arranged in the order of their increasing Lower Excitation Potential (LEP). \\ \vskip2ex
  Only a portion of this table is shown here for guidance regarding its form and content. 
  A machine-readable version of the \\ full table is available.           }
\end{table*}


\end{document}